\documentstyle[12pt]{article}
\begin{document}
\baselineskip18pt
\thispagestyle{empty}
\rightline{FTUAM 94--6}
\rightline{March, 1994}
\vskip1.2cm
\centerline{\large{
\bf{Rigorous QCD Evaluation of Spectrum and}}}
\centerline{\large{
\bf{Other Properties of Heavy
$\bf q\bar{q}$ Systems. }}}
\centerline{\large{
\bf{
II. Bottomium with $n=2$, $l=0,1$
.\footnote{
This work is partially supported by the U.S Department of Energy
and CICYT, Spain.}
}}}
\vskip1.2cm
\centerline{S.~Titard\footnote{Electronic address:
{\tt stephan@nantes.ft.uam.es.}}
 \,and\,F.~J.~Yndur\'{a}in}
\vskip1.2cm
\centerline{{\it Departamento de F\'{\i}sica Te\'orica C-XI}}
\centerline{{\it Universidad Aut\'onoma de Madrid}}
\centerline{{\it 28049 Madrid, Spain}}
\vskip0.8cm
\centerline{\large{Abstract}}
\vskip0.8cm
We calculate the Lamb, fine and hyperfine shifts in $b\bar b$ with
$n=2$, $l=0,1$. Radiative corrections as well as leading
nonperturbative corrections (known to be due to the gluon condensate)
are taken into account. The calculation is parameter-free, as we
take $\Lambda$, ${\langle \alpha_s G^2 \rangle}$
from independent sources. Agreement with
experiment is found at the expected level $\sim 30\%$. Particularly
interesting is a prediction for the hyperfine splitting, $M_{\rm average}
(2^3P)-M(2^1P_1) = 1.7 \pm 0.9\, {\rm MeV}$,
opposite in sign to the $c\bar c$
one ($\approx -0.9\, {\rm MeV}$), and where
the nonzero value of ${\langle \alpha_s G^2 \rangle}$ plays
a leading role.
\setcounter{page}{0}
\newpage
\baselineskip22pt

\section{Introduction}

In a previous paper~\cite{bb:ty} (hereafter to be referred as
TY\footnote{We will freely use the notation of TY}) we presented
an evaluation of the potential for heavy $q\bar q$
systems~\cite{bb:ty,bb:gupt}.
The evaluation included relativistic effects, one--loop
radiative corrections and (for the spin--independent part) the
dominating two--loop ones. With this we evaluated a number
of quantities, taking into account also leading nonperturbative
corrections, which are known~\cite{bb:leut} to be due to the contributions
of the gluon condensate. It was shown that a very good account
could be given of the lowest lying $b\bar b$ bound states
(some features of $c\bar c$ were also discussed). Notably,
both the energy and wave function (this last through $e^+e^-$ decay) of
the states with $n=1$ were given; the splittings between these states
and those with $n=2,\;l=0,1$ were reproduced in what is essentially a zero
parameter calculation using only the known values of the basic QCD
parameters,
\begin{eqnarray}
 \Lambda(n_f=4,{\rm 2\, loops}) &=& 200 \phantom{a}^{+80}_{-60}
\, {\rm MeV}
\nonumber
\\
\label{eq:paras}
 {\langle \alpha_s G^2  \rangle}
&=& 0.042 \pm 0.020 \; {\rm GeV}^4
\\
\nonumber
 m_b &=& 4906 \phantom{a}^{+69}_{-51} \phantom{a}^{-4}_{+4}
\phantom{a}^{+11}_{-40} \;{\rm MeV}.
\end{eqnarray}
Actually we preferred in TY to {\em deduce} $m_b$ from the mass of
the $\Upsilon(1S)$ state. The errors given for this quantity in
(\ref{eq:paras}) correspond to that in $\Lambda$ (the first), to that
in the gluon condensate (the second); the third is an estimated
systematic error.

The value of $m_b$ given in (\ref{eq:paras}) is for the {\em pole}
mass, which is the appropriate quantity to be used in a
Schr{\"o}dinger equation. It corresponds to a running mass value
of
\begin{equation}
\overline{m}_b(\overline{m}_b^2)= 4397
 \phantom{a}^{+7}_{-2}
 \phantom{a}^{-3}_{+4}
 \phantom{a}^{+16}_{-32}
\, {\rm MeV},
\end{equation}
which compares favorably with the SVZ estimate~\cite{bb:shif}
of $ 4250 \pm 100 \,{\rm MeV}$.

For some of the states with $n=2$, $l=1,0$ no result could be given;
only the {\em perturbative} contributions were presented and they failed
to reproduce the experimental values. This was because the
nonperturbative corrections, more involved than for the $n=1$ case, had
not been calculated at the time.

In the present paper we finish the calculation of the leading
nonperturbative (NP, henceforth) contributions to the $n=2$ states.
We are thus able to present a complete, rigorous and parameter--free
QCD evaluation of the full $n=1$ and $n=2$, $l=1,\,0$ bottomium
system. For some of the quantities the NP corrections
(which are always large) are under control; for some others the
calculation loses reliability. By and large, nevertheless, a coherent
picture and good agreement with experiment are obtained.

NP corrections grow very fast with $n$ so for $n\ge 3$ they get
so large (for $b\bar b$) that a QCD calculation based on leading
effects becomes meaningless as was indeed to be expected. However, we
present some results for $n=3,4,5$ with a view to future applications
to the $t\bar t$ system for which NP corrections remain small up to
$n \sim 5$.

This paper is organized as follows: the perturbative $q\bar q$
hamiltonian is reproduced in Sec.~2 for ease of reference.
The NP corrections to the interaction are evaluated in Sec.~3.
Sec.~4 contains the ensuing shifts in energies
and wave functions, which are then applied in Sec.~5 to
the complete evaluation of $n=1,2$, $l=0,1$, $j=0,1,2$ and
spin $s=0,\,1$
bound states of $b\bar b$. The article
is finished in Sec.~6 with numerical
results and Conclusions.

%\newpage

\section{The perturbative QCD Potential.}

We present here the Hamiltonian for the $q\bar q$ system
for ease of reference. We write it separating the
spin--independent, LS, tensor and hyperfine pieces as follows:
\begin{eqnarray}
\label{eq:hsi}
 H_{\rm s.i.} &=& H^{(0)}
- \frac{C_F \beta_0 \alpha_s(\mu^2)}{2 \pi}
 \frac{\ln r\mu}{r},
\\
\label{eq:hzero}
 H^{(0)} &=& - \frac{1}{m}\Delta - \frac{C_F
\widetilde{\alpha}_s(\mu^2)}{r},
\\
\label{eq:alst}
\widetilde{\alpha}_s(\mu^2)
&=& \left[ 1 + \frac{a_1+\gamma_{\rm E}\beta_0/2}{\pi}
\alpha_s(\mu^2) \right] \alpha_s(\mu^2);
\\
\nonumber
V_{\rm LS}(\vec r) &=& \frac{3 C_F \alpha_s(\mu^2)}{2 m^2 r^3}
 \vec L \cdot \vec S
\\
\label{eq:hls}
 && \times \left\{ 1 + \left[
 \frac{\beta_0}{2}(\ln r \mu - 1) +2 (1-\ln m r)
 +\frac{125 - 10\,n_f}{36} \right] \frac{\alpha_s}{\pi} \right\}
\\
\nonumber
V_{\rm T}(\vec r)
&=& \frac{C_F \alpha_s(\mu^2)}{4 m^2 r^3} S_{12}(\vec r)
\\
\label{eq:ht}
&& \times \left\{
 1 + \left[
 D +\frac{\beta_0}{2} \ln r \mu - 3\ln m r \right]
  \frac{\alpha_s}{\pi}  \right\}
\\
\label{eq:hhf}
V_{\rm hf}(\vec r)
&=& \frac{4 \pi C_F \alpha_s(\mu^2)}{3 m^2} {\vec S}^2
 \left\{ \phantom{\frac{a}{a}}\!\!\! \delta(\vec r)  \right.
\\
\nonumber
 && \left. + \left[ \frac{\beta_0}{2}
\left(
\frac{1}{4 \pi} {\rm reg}\frac{1}{r^3}
+ (\ln \mu)\delta(\vec r)
\right)
- \frac{21}{4}
\left(
\frac{1}{4 \pi} {\rm reg}\frac{1}{r^3}
+(\ln m + B)\delta(\vec r)
\right)
\right]
 \frac{\alpha_s}{\pi}
\right\} \;.
\end{eqnarray}
Here,
\begin{eqnarray}
\nonumber
 && C_A = 3, \, T_F = 1/2, \,\beta_0 = 11 - \frac{2\, n_f}{3},
\,\beta_1 = 102 - \frac{38\, n_f}{3}
\\
\nonumber
 && a_1 = \frac{31\, C_A - 20 \,T_F\, n_f}{36} ,
\\
\nonumber
 && B = \frac{3}{2}(1 -\ln 2)\, T_F - \frac{5}{9}\, T_F\, n_f
+\frac{11\, C_A - 9\, C_F}{18} ,
\\
\nonumber
 && D = \frac{4}{3} \left(3 - \frac{\beta_0}{2} \right)
 + \frac{65}{12} - \frac{5\, n_f}{18}.
\end{eqnarray}
$ \vec S = {\vec S}_1 + {\vec S}_2$ is the total spin, $\vec L$ the
orbital angular momentum and
\begin{displaymath}
\nonumber
S_{12}(\vec r) = 2 \sum_{i j} S_i S_j \left(
 \frac{3}{r^2} r_i r_j - \delta_{ij} \right) \,.
\end{displaymath}
$n_f$ is the number of {\em active} flavours. The running coupling
constant we take to two loops,
\begin{displaymath}
\alpha_s(\mu^2) = \frac{4 \pi}{\beta_0 \ln \mu^2/\Lambda^2}
 \left\{
1 - \frac{\beta_1}{\beta_0^2}
\frac{ \ln\ln \mu^2/\Lambda^2}{\ln \mu^2 / \Lambda^2}
\right\}\, .
\end{displaymath}
We have lumped the constant piece of the one--loop correction into
$\widetilde{\alpha}_s$
(Eq.~(\ref{eq:alst})) because the ensuing potential is
still Coulombic and therefore $H^{(0)}$ may still be solved exactly.
The relativistic, full one loop and leading two loop corrections to
the spin--independent piece
are known; see TY for details. We will not need them now.
The total Hamiltonian is of course
\begin{equation}
\label{eq:hp}
 H_{\rm p} = H_{\rm s.i.} + V_{\rm LS} + V_{\rm T}
 + V_{\rm hf} \, ,
\end{equation}
where the index p emphasizes that only {\em perturbative}
contributions are taken into account.

A result that we take over from TY is the form of the
(spin--independent) wave functions
$ {\bar \Psi}^{(0)}_{nl}$
pertaining to the Hamiltonian $H_{\rm s.i.}$. They are easiest
obtained with a variational method; one finds that they are given
by a formula like that for the wave functions of the Coulombic
Hamiltonian $H^{(0)}$ with the replacement of the "Bohr radius",
\begin{displaymath}
 a = \frac{2}{m C_F \widetilde{\alpha}_s} \, ,
\end{displaymath}
by
\begin{eqnarray}
\label{eq:bnl}
 b(n,l) &=& a \left\{
1 + \frac{\ln(n \mu / m C_F \widetilde{\alpha}_s)
+ \psi(n+l+1) - 1}{2 \pi}
 \, \beta_0 \alpha_s \right\}^{-1} \, ,
\\
\nonumber
{\bar \Psi}^{(0)}_{nl}(\vec r) &=& \Psi^{(0)}_{nl}(\vec r; a
\rightarrow b) \, .
\end{eqnarray}
A few explicit expressions may be found in Appendix~II.

In particular the wave function at the origin becomes
\begin{eqnarray}
\nonumber
\Psi^{(0)}_{nl}(0)
&\rightarrow& {\bar \Psi}^{(0)}_{nl}(0) =
 \{ 1 + \delta_{\rm wf}(n,l) \} \Psi^{(0)}_{nl}(0) \, ,
\\
\label{eq:dwf}
\delta_{\rm wf}(n,l) &=& \frac{3 \beta_0}{4 \pi}
 \left[ \ln(\frac{n \mu}{m C_F \widetilde{\alpha}_s})
+ \psi(n+l+1) -1 \right]
\alpha_s \,.
\end{eqnarray}
As stated, $\Psi^{(0)}_{nl}$ is the solution of the equation
\begin{equation}
\label{eq:unperta}
 H^{(0)} \Psi^{(0)}_{nl} = E_n^{(0)} \Psi^{(0)}_{nl} \, ,
\; E_n^{(0)} = - \frac{(C_F \widetilde{\alpha}_s)^2}{4 n^2} m \,.
\end{equation}
When taking into account the full $H_{\rm s.i.}$ the energies are
shifted to $ {\bar E}^{(0)}_{nl} $,
\begin{equation}
\label{eq:ebarnl}
{\bar E}^{(0)}_{nl} = E_{nl}^{(0)} - \frac{C_F \beta_0 \,\alpha_s^2
\,\widetilde{\alpha}_s}
{4 \pi n^2} \left\{
 \ln\frac{n \mu}{m C_F \alpha_s} + \psi(n+l+1) \right\} m \, .
\end{equation}

A last word about the notation: the superindex $(0)$ in say,
$ \Psi^{(0)}$, $E^{(0)}$ means "of zero order with respect
to {\em nonperturbative} (NP) effects".

\section{ The Nonperturbative Interactions.}

It can be shown (TY and \cite{bb:leut,bb:shif,bb:yndu})
that the leading
NP interactions, {\em at short distances}, are those
associated with the gluon condensate; and, of these, the dominant
ones are those where two gluons are attached to the quarks. These
interactions are equivalent, in the nonrelativistic limit (including
first order relativistic corrections) to those obtained assuming
the quarks to move inside a medium of constant, random chromoelectric,
$\vec {\cal E}$ and chromomagnetic, $\vec {\cal B}$ fields. Because
the fields are constant they may be considered to be classical; and
because they are random we may take them of zero average value
\begin{displaymath}
 \langle \vec {\cal E}\, \rangle \,=\,
 \langle \vec {\cal B} \, \rangle \,=\, 0 \;.
\end{displaymath}
The average is taken in the physical vacuum. Quadratic averages
are non--vanishing and may be related to the gluon condensate.
With $i,\, j$ spatial indices and $a, \, b$ color ones one has
(for $N_c =3$ colors)
\begin{equation}
\label{eq:quadav}
 \langle g^2 {\cal B}^i_a {\cal B}^j_b  \rangle =
 - \langle g^2 {\cal E}^i_a {\cal E}^j_b  \rangle =
\frac{\pi \delta_{ij} \delta_{ab}}{3 (N_c^2-1)}
 \langle \alpha_s G^2  \rangle  \;.
\end{equation}

The relativistic interaction of a quark (labeled with index $1$)
with classical vector fields may be described by the Dirac Hamiltonian
\begin{equation}
\label{eq:dirac}
 H_{D1}= i {\vec \alpha}_1 \cdot {\vec \nabla}_1
 -g \gamma \cdot
{A\mkern -7.5mu{\lower1.6ex\hbox{$\widetilde {}$}}\;\;}
({\vec r}_1) + \beta_1 m \, ,
\end{equation}
${A^\mu\mkern -14mu{\lower1.6ex\hbox{$\widetilde {}$}}\;\;\,\,\,}
 = \sum\limits_a \,
{t^a\mkern -10mu{\lower1.6ex\hbox{$\widetilde {}$}}\;\;}
 A^\mu_a $ being gluon fields (in matrix notation). A convenient
gauge is that in which
\begin{displaymath}
{A^0_1\mkern -13.5mu{\lower1.6ex\hbox{$\widetilde {}$}}\;\;\,}
 = - {\vec r}_1 \cdot
{{\vec {\cal E}}\mkern -7mu{\lower1.6ex\hbox{$\widetilde {}$}}\;\,}
\,,\,
{{\vec A}_1\mkern -13.5mu{\lower1.6ex\hbox{$\widetilde {}$}}\;\;\,}
 = - \frac{1}{2} {\vec r}_1 \times
{{\vec {\cal B}}\mkern -7mu{\lower1.6ex\hbox{$\widetilde {}$}}\;\,}
\;.
\end{displaymath}
To solve our problem
one can apply a Foldy--Wouthuysen transformation~\cite{bb:bj} to
obtain the Hamiltonian (correct including first order
relativistic effects)
\begin{eqnarray}
\label{eq:hfw}
 H_{FW1} &=& m + \frac{1}{2 m}
({\vec p}_1 -g
{{\vec A}_1\mkern -13.5mu{\lower1.6ex\hbox{$\widetilde {}$}}\;\;\,}
 )^2 - \frac{1}{8 m^3} {{\vec p}_1}^{\;4}
\\
\nonumber
&&
 - \frac{g}{m} {\vec S}_1 \cdot
{{\vec {\cal B}}\mkern -7mu{\lower1.6ex\hbox{$\widetilde {}$}}\;\,}
 - \frac{g}{2 m^2} {\vec S}_1 \cdot (
{{\vec {\cal E}}\mkern -7mu{\lower1.6ex\hbox{$\widetilde {}$}}\;\,}
 \times {\vec p}_1 ) \, ,
\end{eqnarray}
${\vec S}_1$ the spin operator and ${\vec p}_1 = - i {\vec \nabla}_1$.
Adding to this the Hamiltonian of the antiquark ($g \rightarrow -g,
\, {\vec r}_1 \rightarrow {\vec r}_2$) and their interactions given
in the previous section we find the full hamiltonian, which
now includes leading NP effects,
\begin{eqnarray}
\label{eq:fullham}
 H &=& H^{(0)} - \frac{C_F \beta_0 \alpha_s^2}{2 \pi} \frac{\ln r \mu}
{r} + V_{\rm LS} + V_{\rm T} + V_{\rm hf}
\\
\nonumber
 &&
 - g  {\vec r} \cdot
{{\vec {\cal E}}\mkern -7mu{\lower1.6ex\hbox{$\widetilde {}$}}\;\,}
 + \frac{g}{2 m^2} ({\vec S}\times {\vec p})  \cdot
{{\vec {\cal E}}\mkern -7mu{\lower1.6ex\hbox{$\widetilde {}$}}\;\,}
 - \frac{g}{m} ({\vec S}_1 - {\vec S}_2)  \cdot
{{\vec {\cal B}}\mkern -7mu{\lower1.6ex\hbox{$\widetilde {}$}}\;\,}
 \,.
\end{eqnarray}
$H^{(0)}$, $V_{\rm LS}$, $V_{\rm T}$, $V_{\rm hf}$ are given by
Eqs.~(\ref{eq:hzero}) to (\ref{eq:hhf}). Some of the
peculiarities of Eq.~(\ref{eq:fullham}), in particular the
absence of an ${\vec L}\cdot {\vec S}$ interaction as well as
the presence of a term involving the differences of the spins, had
been noted in the similar case of the Zeeman effect in
positronium~\cite{bb:akhi}. In Eq.~(\ref{eq:fullham}) we have omitted
a term obtained when expanding the square
$({\vec p}_1 -g
{{\vec A}_1\mkern -13.5mu{\lower1.6ex\hbox{$\widetilde {}$}}\;\;\,}
 )^2
$
in Eq.~(\ref{eq:hfw}), viz., the piece
$
{{\vec A}_1^2\mkern -13.5mu{\lower1.6ex\hbox{$\widetilde {}$}}\;\;\,}
\,$.
It would have produced a term $\pi
{\langle \alpha_s G^2  \rangle}\,r^2\,/(48 N_c m)$, to be added to
Eq.~(\ref{eq:fullham}). The reason for its omission is that
it gives {\em subleading} corrections to all processes
(as compared to the contributions of the other terms).

Before embarking upon detailed calculations, let us elaborate on this
matter of leading and subleading corrections. Because
\begin{eqnarray}
\nonumber
 \langle r \rangle &\sim& a \;=\; \frac{2}{m C_F
{\widetilde{\alpha}_s}} \, ,
\\
\nonumber
\langle \, p\, \rangle &\sim& m \,v \;\sim\; m C_F
{\widetilde{\alpha}_s}\, ,
\end{eqnarray}
it follows that the NP terms in Eq.~(\ref{eq:fullham}) are
\begin{eqnarray}
\label{eq:orders}
&& - g  {\vec r} \cdot
{{\vec {\cal E}}\mkern -7mu{\lower1.6ex\hbox{$\widetilde {}$}}\;\,}
\;\sim\; \frac{1}{{\widetilde{\alpha}_s}} \;\; , \;\;
\frac{g}{2 m^2} ({\vec S}\times {\vec p})  \cdot
{{\vec {\cal E}}\mkern -7mu{\lower1.6ex\hbox{$\widetilde {}$}}\;\,}
\;\sim\; {\widetilde{\alpha}_s}\, ,
\\
\nonumber
&& - \frac{g}{m} ({\vec S}_1 - {\vec S}_2)  \cdot
{{\vec {\cal B}}\mkern -7mu{\lower1.6ex\hbox{$\widetilde {}$}}\;\,}
\;\sim\; ({\widetilde{\alpha}_s})^0 \, .
\end{eqnarray}
This simplifies enormously the calculation at the leading order
as seldom more than one, and at most two terms need to be considered.
A further simplification is that, with the only exception of the
hyperfine splitting for $n=2$, $l=1$, only the tree level piece
of $H_{\rm p}$ has to be taken into account when evaluating leading
NP effects.

\section{Energy and Wave Function Shifts.}

\subsection{Spin--independent Shifts.}

Although most of the spin--independent shifts of
energies and wave functions were discussed in TY
and~\cite{bb:leut},
we give here a detailed calculation for ease of reference,
to correct an error common to TY
and~\cite{bb:leut}, to present the results for the $n=2$
wave functions and to explain in this simple case the way the
calculation works.

The effects of the nonzero condensate are evaluated with the help
of perturbation theory. The perturbation consists of the terms
(cf. Eq.~(\ref{eq:fullham})),
$ - g  {\vec r} \cdot
{{\vec {\cal E}}\mkern -7mu{\lower1.6ex\hbox{$\widetilde {}$}}\;\,}
$,
$\displaystyle
 \frac{g}{2 m^2} ({\vec S}\times {\vec p})  \cdot
{{\vec {\cal E}}\mkern -7mu{\lower1.6ex\hbox{$\widetilde {}$}}\;\,}
$,
$ \displaystyle
- \frac{g}{m} ({\vec S}_1 - {\vec S}_2)  \cdot
{{\vec {\cal B}}\mkern -7mu{\lower1.6ex\hbox{$\widetilde {}$}}\;\,}
$.
Because, for spin independent effects, the first term gives a nonzero
result we may neglect the others which would contribute corrections
of higher order in $\alpha_s$, cf. Eq.~(\ref{eq:orders}). Second order
perturbation theory is required as only quadratic terms in
$
{{\vec {\cal E}}\mkern -7mu{\lower1.6ex\hbox{$\widetilde {}$}}\;\,}
$ will give a nonvanishing contribution, as discussed in the previous
section, Eq.~(\ref{eq:quadav}) and above. The method of evaluation,
for this particular case, has been developed by Leutwyler,
Ref.~\cite{bb:leut}, and is related to Kotani's treatment of the
second order Stark effect~\cite{bb:kota}, up to color and
angular momentum complications that we now discuss.

We denote the solutions of the unperturbed Hamiltonian by
\begin{eqnarray}
\nonumber
&& H^{(0)}
\,\Bigl\vert \Psi^{(0)}_{nlM} \Bigr\rangle
\,=\, E_n^{(0)}
\,\Bigl\vert \Psi^{(0)}_{nlM} \Bigr\rangle \,,
\\
\nonumber
&& E_n^{(0)} \,=\, - \frac{1}{m\, a^2\, n^2} \,=\, - \frac{C_F^2\,
{\widetilde{\alpha}_s}}{4\, n^2} \,m \,,
\\
\label{eq:unpertb}
&&\Psi^{(0)}_{nlM} \,=\, Y^l_M({\vec r}/r) \,R^{(0)}_{nl}(r) \;.
\end{eqnarray}
(We have omitted the trivial rest mass energy term).
The $R^{(0)}_{nl}(r)$ are identical to the standard Coulombic wave
functions for the hydrogen atom with the replacement of the "Bohr
radius" by $\displaystyle a=\frac{2}{m C_F {\widetilde{\alpha}_s}}$.
Second order perturbation theory yields immediately the energy and
wave function shifts:
\begin{displaymath}
E = E^{(0)} + E^{\rm NP} \;;\;
\Psi = \Psi^{(0)} + \Psi^{\rm NP}
\end{displaymath}
with
\begin{equation}
\label{eq:energy}
E^{\rm NP}_{nl} = -
 \sum_{ij,\,ab}
\,\left\langle \Psi^{(0)}_{nlM} \left\vert
 g \,r_i\, {\cal E}^i_a\, t^a \,
\frac{1}{H^{(0)}-E_n^{(0)}} \,
 g\, r_j\, {\cal E}^j_b\, t^b\,
\right\vert \Psi^{(0)}_{nlM} \right\rangle
\end{equation}
and
\begin{equation}
\label{eq:wfn}
\Bigl\vert \Psi^{\rm NP}_{nlM} \Bigr\rangle
 \,=\,  \sum_{ij,\,ab} \,P_{nl}
\,\frac{1}{H^{(0)}-E_n^{(0)}} \,P_{nl}\,\,
 g\, r_i\, {\cal E}^i_a\, t^a
\,\frac{1}{H^{(0)}-E_n^{(0)}} \,
 g\, r_j \,{\cal E}^j_b\, t^b\,
\Bigl\vert \Psi^{(0)}_{nlM} \Bigr\rangle \;.
\end{equation}
\newpage
\noindent Here
\begin{displaymath}
 P_{nl} \,=\, 1 - \Bigl\vert\Psi^{(0)}_{nlM} \Bigr\rangle
\Bigl\langle \Psi^{(0)}_{nlM} \Bigr\vert
\end{displaymath}
is the projector orthogonal to the $nl$ state. It does not appear
in Eq.~(\ref{eq:energy}) because
\begin{displaymath}
\langle \Psi^{(0)}_{nlM} \vert
 {\vec r} \cdot
{{\vec {\cal E}}\mkern -7mu{\lower1.6ex\hbox{$\widetilde {}$}}\;\,}
\vert \Psi^{(0)}_{nlM} \rangle \;=\; 0 \;.
\end{displaymath}
The expressions~(\ref{eq:energy}), (\ref{eq:wfn}) are first simplified
by replacing
\begin{equation}
\label{eq:redE}
 g {\cal E}_a^i \ldots g {\cal E}_b^j
 \rightarrow
 - \frac{\delta_{ij} \delta_{ab}}{24} \pi
{\langle \alpha_s G^2  \rangle} \,,
\end{equation}
recall Eq.~(\ref{eq:quadav}).

Next we take care of the color algebra. The one--gluon exchange
potential is given, when acting on arbitrary color states by
\begin{equation}
 - \frac{\alpha_s}{r}\, \sum_a \, t^a_{i i'} t^b_{k' k}\;.
\end{equation}
If the initial (and final) states are color singlets we may average
\begin{displaymath}
 \frac{1}{\sqrt{N_c}} \,
\sum_{ik} \delta_{ik}
 \frac{1}{\sqrt{N_c}} \,
\sum_{i'k'} \delta_{i'k'} \,,
\end{displaymath}
and then we get the potential, and Hamiltonian,
\begin{displaymath}
 - \frac{C_F
{\widetilde{\alpha}_s}}{r} \;, \;
 H^{(0)} =  - \frac{1}{m} \Delta
 - \frac{C_F
{\widetilde{\alpha}_s}}{r} \,; \,
\end{displaymath}
we have incorporated, as we always do everywhere, the Coulombic
piece of the one--loop corrections into
${\widetilde{\alpha}_s}$.

In Eqs.~(\ref{eq:energy}) and (\ref{eq:wfn}), however, the states
$\Bigl\vert \Psi^{(0)}_{nlM} \Bigr\rangle$ are
certainly color singlets: hence
the matrices $t^b$ (for example) when acting on them will produce
a {\em color octet} state. For a color octet the potential and
Hamiltonian are
\begin{equation}
\label{eq:hoctet}
 \frac{
{\widetilde{\alpha}_s}}{2 N_c r} \;, \;
 H^{\,'(0)} =  - \frac{1}{m} \Delta
 + \frac{
{\widetilde{\alpha}_s}}{2 N_c r} \,.
\end{equation}
One then finds
\begin{equation}
\label{eq:coltrick}
 \sum_{ab} \, \delta_{ab} t^a
\,\frac{1}{H^{(0)}-E_n^{(0)}} \,
 t^b \,\Bigl\vert {\rm singlet} \Bigr\rangle =
 C_F \,\frac{1}{H^{\,'(0)}-E_n^{(0)}} \,
\,\Bigl\vert {\rm singlet} \Bigr\rangle \;.
\end{equation}
\newpage
Putting this together with Eq.~(\ref{eq:redE}) into Eqs.~(
\ref{eq:energy}) and (\ref{eq:wfn}) gives the formulas
\begin{eqnarray}
\label{eq:energya}
E^{\rm NP}_{nl}
 &=&  \frac{\pi {\langle \alpha_s G^2  \rangle}}{6 N_c}
 \;\sum_{i}
\left\langle \Psi^{(0)}_{nlM} \left\vert
r_i \,\frac{1}{H^{\,'(0)}-E_n^{(0)}} \, r_i
\right\vert \Psi^{(0)}_{nlM} \right\rangle \,,
\\
\nonumber
\Bigl\vert \Psi^{\rm NP}_{nlM} \Bigl\rangle
 &=&  - \frac{\pi {\langle \alpha_s G^2  \rangle}}{6 N_c}
\; P_{nl} \,\frac{1}{H^{(0)}-E_n^{(0)}} \,P_{nl}
\\
\label{eq:wfna}
 && \times \;
\sum_{i}
 \,r_i  \,\frac{1}{H^{\,'(0)}-E_n^{(0)}} \, r_i \,
\Bigl\vert \Psi^{(0)}_{nlM} \Bigr\rangle \;,
\end{eqnarray}
which takes care of color complications, so we turn to deal with
angular momentum. Obviously the perturbation is
rotationally invariant so the third component of angular
momentum, $M$, is not affected by it; but the total angular
momentum algebra is not entirely trivial. We write
\begin{displaymath}
\sum_{i}
 r_i  \,\frac{1}{H^{\,'(0)}-E_n^{(0)}} \, r_i
 \,=\,
\sum_{\lambda}
 r^*_\lambda  \,\frac{1}{H^{\,'(0)}-E_n^{(0)}} \, r_\lambda
\,,
\end{displaymath}
where $\lambda=0,\,\pm1$, and the $r_\lambda$'s are
spherical components,
\begin{displaymath}
 r_{\pm 1} = \mp \frac{1}{\sqrt{2}} (r_1 \pm i r_2)
\;,\; r_0 = r_3 \;.
\end{displaymath}
Using the formulas
\begin{equation}
\label{eq:spher}
\frac{1}{r} \,r_\lambda \,=\, \sqrt{\frac{4 \pi}{3}}
\,Y^1_\lambda({\vec r}/r) \,\;;\;\,
\frac{1}{r} \,r^*_\lambda \,=\,
 (-1)^{\lambda}\,
\sqrt{\frac{4 \pi}{3}}
\,Y^{\,1}_{-\lambda}({\vec r}/r) \;;
\end{equation}
and the addition theorem for spherical harmonics we get
\begin{eqnarray}
\nonumber
 r_\lambda Y^l_M \,=\, r \!
\sum_{l'= \vert l-1 \vert,\vert l+1 \vert}
 C_M(l,l',\lambda) \,Y^{l'}_{M+\lambda}\,,
\\
\nonumber
 C_M(l,l',\lambda) \,=\, \sqrt{\frac{2 l +1}{2 l' +1}}
 \;(l,M;1,\lambda \vert l')\,(l,0;1,0 \vert l')
\end{eqnarray}
with $(\ldots\vert\ldots)$ the standard Clebsch--Gordan
coefficients.

When acting on a function with well--defined
angular momentum $l$ we have
\begin{eqnarray}
\nonumber
\,\frac{1}{H^{\,'(0)}-E_n^{(0)}} \,
\Bigl\vert \,l\, \Bigr\rangle
\,&=& \,
\,\frac{1}{H^{\,'(0)}_l-E_n^{(0)}} \,
\Bigl\vert \,l\, \Bigr\rangle \,,
\\
\label{eq:angtrick}
\,\frac{1}{H^{(0)}-E_n^{(0)}} \,
\Bigl\vert \,l\, \Bigr\rangle
\,&=& \,
\,\frac{1}{H^{(0)}_l-E_n^{(0)}} \,
\Bigl\vert \,l\, \Bigr\rangle \,,
\end{eqnarray}
where
\begin{equation}
\label{eq:hl}
H_l \,=\, -\frac{1}{m} \frac{1}{r^2}
\frac{\partial}{\partial r}
\left( r^2 \frac{\partial}{\partial r} \right) +
\frac{l(l+1)}{m r^2} + \frac{\kappa {\widetilde{\alpha}_s}}{r}
\,,
\end{equation}
with $\kappa= -C_F$ for $H_l^{(0)}$ and
$\kappa= 1/(2\, N_c)$ for $H_l^{\,'(0)}$.
Using this and the explicit values of the Clebsch--Gordan
coefficients we find that Eqs.~(\ref{eq:energya}),
(\ref{eq:wfna}) become
\begin{eqnarray}
\nonumber
E^{\rm NP}_{nl} =
 &=&  \frac{\pi {\langle \alpha_s G^2  \rangle}}{6 N_c}
 \, \frac{1}{2 l +1}
\\
\label{eq:energyb}
&& \times \,\left\langle R^{(0)}_{nl} \left\vert
\,r \left\{
\frac{l}{H_{l-1}^{\,'(0)}-E_n^{(0)}} \,
+ \,\frac{l+1}{H_{l+1}^{\,'(0)}-E_n^{(0)}}
\right\}  \,r
\right\vert R^{(0)}_{nl} \right\rangle \,,
\\
\nonumber
\Bigl\vert R^{\rm NP}_{nl} \Bigl\rangle
 &=& - \frac{\pi {\langle \alpha_s G^2  \rangle}}{6 N_c}
 \, \frac{1}{2 l +1}
\; P_{nl} \,\frac{1}{H_l^{(0)}-E_n^{(0)}} \,P_{nl}
\\
\label{eq:wfnb}
 && \times \;
\,r\, \left\{
\frac{l}{H_{l-1}^{\,'(0)}-E_n^{(0)}} \,
+ \,\frac{l+1}{H_{l+1}^{\,'(0)}-E_n^{(0)}}
\right\} \, r\,
\Bigl\vert R^{(0)}_{nl} \Bigr\rangle \;.
\end{eqnarray}
We have succeeded in separating the color and angular
variables to obtain equations involving only the radial variable
and radial wave functions. To finish the calculations all that is
needed is to find the inverses
\begin{displaymath}
\frac{1}{H_l^{(0)}-E_n^{(0)}}\ldots R^{(0)}_{nl}
\;,\;
\,\frac{1}{H_l^{\,'(0)}-E_n^{(0)}}\ldots R^{(0)}_{nl} \,.
\end{displaymath}
This is described in Appendix~I. The ensuing expressions
for the $E_{nl}^{\rm NP}$ and $R^{\rm NP}_{nl}$ are collected
in Appendix~II for a few values of $n,\,l$ and will be
employed later on. The expression we get for $E_{nl}^{\rm NP}$
agrees with that found by Leutwyler~\cite{bb:leut} and also
$R^{\rm NP}_{10}$, the only wave function calculated in
Ref.~\cite{bb:leut}, agrees with our evaluation.

We have not succeeded in obtaining a closed general formula
for $R^{\rm NP}_{nl}$ (for $E^{\rm NP}_{nl}$ one is given in
Ref.~\cite{bb:leut}) but a few general properties may be
inferred from Eqs.~(\ref{eq:energyb}), (\ref{eq:wfnb}).
Because
\begin{displaymath}
 \langle r \rangle_{nl} \,=\,
\frac{a}{2} (3 n^2 - l(l+1)) \, \sim \,
\frac{n^3}{{\widetilde{\alpha}_s}}
\end{displaymath}
and each energy denominator yields a factor $\displaystyle
\frac{1}{{\widetilde{\alpha}_s} n^2}$ (see Appendix~I) we expect
\begin{displaymath}
 E^{\rm NP}_{nl} \,\sim\, \frac{n^6}{{\widetilde{\alpha}_s}^4}
\; , \;
 R^{\rm NP}_{nl} \,\sim\, \frac{n^8}{{\widetilde{\alpha}_s}^6}
\;.
\end{displaymath}
It thus follows that the importance of nonperturbative effects
grows very rapidly with $n$. Moreover we expect them to be smaller
for energies than for wave functions and, generally, to be larger
when $l=0$ than for $l\neq 0$ (for the same value of $n$). These
properties may be verified explicitly in the expressions collected
in Appendix~II.

The energies and wave functions correct to leading order in NP effects
and including one--loop corrections are then
\begin{eqnarray}
\nonumber
 E_{nl} &=& {\bar E}^{(0)}_{nl} + E^{\rm NP}_{nl} \,,
\\
\label{eq:fullenwf}
R_{nl}(r) &=& {\bar R}^{(0)}_{nl}(r) + R^{\rm NP}_{nl}(r)\,,
\\
\Psi_{nlM} &=& Y^l_M({\vec r}/r)\,R_{nl}(r)\,,
\end{eqnarray}
the ${\bar R}^{(0)}$, ${\bar E}^{(0)}$ being as given
in Eqs.~(\ref{eq:bnl}), (\ref{eq:dwf}).

\subsection{Hyperfine splittings}

The hyperfine splittings are caused by the interactions that depend
only on spin; they are $V_{\rm hf}$ in Eq.~(\ref{eq:hhf}), and
the piece
\begin{displaymath}
 - \frac{g}{m} ({\vec S}_1 - {\vec S}_2)  \cdot
{{\vec {\cal B}}\mkern -7mu{\lower1.6ex\hbox{$\widetilde {}$}}\;\,}
\end{displaymath}
in Eq.~(\ref{eq:fullham}). Besides the splitting caused directly by
the last term, there is a nonperturbative contribution
indirectly generated by
$ - g {\vec r} \cdot
{{\vec {\cal E}}\mkern -7mu{\lower1.6ex\hbox{$\widetilde {}$}}\;\,}
$. This contribution that we will call "internal", comes about because,
when evaluating the expectation values
\begin{displaymath}
 \langle \Psi \vert
 \, V_{\rm hf} \,
 \vert \Psi \rangle
\end{displaymath}
we should use the wave function including the NP corrections
discussed in the previous Subsection:
\begin{eqnarray}
\nonumber
&& \langle \Psi_{nl} \vert
 \, V_{\rm hf} \,
 \vert \Psi_{nl} \rangle
=
 \langle {\bar \Psi}^{(0)}_{nl}+ \Psi_{nl}^{\rm NP} \vert
 \, V_{\rm hf} \,
 \vert {\bar \Psi}^{(0)}_{nl}+ \Psi_{nl}^{\rm NP} \rangle
\\
\label{eq:int}
&&
\simeq \langle {\bar \Psi}^{(0)}_{nl} \vert
 \, V_{\rm hf} \,
 \vert {\bar \Psi}^{(0)}_{nl} \rangle
+
 2 \,\langle \Psi^{(0)}_{nl} \vert
 \, V_{\rm hf} \,
 \vert \Psi_{nl}^{\rm NP} \rangle \;.
\end{eqnarray}
The "internal" NP splitting is the last term in Eq.~(\ref{eq:int}):
\begin{equation}
\label{eq:dhfint}
\Delta^{\rm in}_{\rm hf}\,E_{nls} \,=\,
 2 \,\langle \Psi^{(0)}_{nl;s} \vert
 \, V_{\rm hf} \,
 \vert \Psi_{nl;s}^{\rm NP} \rangle
\,.
\end{equation}
To evaluate this to leading order we use the expression
\begin{displaymath}
V_{\rm hf} \simeq \frac{4 \pi C_F \alpha_s}{3 m^2}
 \,\delta({\vec r}) {\vec S}^{\,2}
\end{displaymath}
and thus we get
\begin{equation}
\label{eq:dhfinta}
\Delta^{\rm in}_{\rm hf}\,E_{n0s} \,=\,
 2 s(s+1) \frac{4 \pi C_F \alpha_s}{3 m^2} \frac{1}{4 \pi}
 R^{(0)}_{n0}(0)\,R^{\rm NP}_{n0}(0) \,.
\end{equation}
For $l \neq 0$ the leading piece of $V_{\rm hf}$ gives zero, because
$R^{(0)}_{nl}(0)$ vanishes. We have to take into account
the radiative correction to $V_{\rm hf}$ and then
\begin{eqnarray}
\nonumber
\Delta^{\rm in}_{\rm hf}\,E_{nls} &=&
 2 s(s+1) \frac{4 \pi C_F \alpha_s}{3 m^2}
\left(\frac{\beta_0}{2}-\frac{21}{4}\right)
\frac{1}{4 \pi} \frac{\alpha_s}{\pi}
\\
\label{eq:dhfintb}
&& \times \,\int_0^\infty \!dr\, r^2
 R^{(0)}_{nl}(r)\, \frac{1}{r^3} \,R^{\rm NP}_{nl}(r)
 \;,\;l\neq 0 \;.
\end{eqnarray}
It will turn out that, for $l\neq 0$, this internal shift
will be subleading. This fact is very interesting because this is
one a the few cases where a rigorous QCD analysis yields results
{\em qualitatively} different from the
calculations based on phenomenological
potentials. This we will discuss in detail elsewhere.

The contribution to hyperfine splitting of the interation
$\displaystyle - \frac{g}{m} ({\vec S}_1 - {\vec S}_2)  \cdot
{{\vec {\cal B}}\mkern -7mu{\lower1.6ex\hbox{$\widetilde {}$}}\;\,}
$
we will call "external"~\footnote{In the case of hyperfine
splittings the internal contribution is chromoelectric and the
external one chromomagnetic, but this is not true in other
splittings.}.
It may be calculated as we calculated $E_{nl}^{\rm NP}$ in the
previous Subsection. We find,
\begin{equation}
\label{eq:dhfext}
\Delta^{\rm ex}_{\rm hf}\,E_{nls} \,=\,
 [s(s+1)-3]\, \frac{\pi {\langle \alpha_s G^2  \rangle}}
{6 N_c m^2}
\,\left\langle R^{(0)}_{nl} \left\vert
\,\frac{1}{H_{l}^{\,'(0)}-E_n^{(0)}} \,
 \right\vert R^{(0)}_{nl} \right\rangle
\;.
\end{equation}
The inverse is obtained with the formulas of Appendix~I.

To the NP contributions we have to add tree level (relativistic)
and radiative ones, that we collectively label perturbative: from
TY,
\begin{eqnarray}
\nonumber
\Delta^{\rm p}_{\rm hf}\,E_{n0s} \,&=&\,
 \frac{s(s+1)}{2}\, \frac{C_F^4 \alpha_s(\mu^2)
 {\widetilde{\alpha}_s}^3(\mu^2)}{3 n^3}\,m
[1 + \delta_{\rm wf}(n,0)]^2
\\
\nonumber
&& \times
\left\{
 1 + \left[ \frac{\beta_0}{2}\left(
 \ln\frac{n \mu}{m C_F {\widetilde{\alpha}_s}}
 - \sum_1^n \frac{1}{k} - 1 + \gamma_{\rm E} - \frac{n-1}{2 n}
 \right)
\right. \right.
\\
\label{eq:dhfp}
&& \left. \left.
\qquad
 -\frac{21}{4}\left(
 \ln\frac{n}{C_F {\widetilde{\alpha}_s}}
 - \sum_1^n \frac{1}{k} - \frac{n-1}{2 n} \right) + B \right]
\,\frac{\alpha_s}{\pi} \right\} \;,
\\
\nonumber
\Delta^{\rm p}_{\rm hf}\,E_{nls} \,&=&\,
 \frac{s(s+1)}{2}\, \frac{C_F^4 \alpha_s^2
 {\widetilde{\alpha}_s}^3}{6 \pi n^3 l(l+1)(2 l +1)}
\left( \frac{\beta_0}{2}-\frac{21}{4}\right)
\,m \;, \;\; l\neq 0\,.
\end{eqnarray}
The constants are as in Eq.(\ref{eq:hhf}). The full splitting is
thus
\begin{equation}
\label{eq:dhffull}
\Delta_{\rm hf}\,E_{nls} \,=\,
\Delta^{\rm p}_{\rm hf}\,E_{nls} +
\Delta^{\rm in}_{\rm hf}\,E_{nls} +
\Delta^{\rm ex}_{\rm hf}\,E_{nls} \;,
\end{equation}
with the various pieces given in Eqs.~(\ref{eq:dhfint}) to
(\ref{eq:dhfp}).

\subsection{Fine splittings}

Also here we have "internal" and "external" contributions.
The internal ones are, as before, induced by the NP modification
of the wave function. The calculation is somewhat complicated
because now two operators, the LS and Tensor ones
(Eqs.~(\ref{eq:hls}) and (\ref{eq:ht}) ) contribute.
We find\footnote{We consider that the states correspond to total
spin $s=1$. For $s=0$, $\Delta^{\rm in}_f E^{\,s=0}_{nlj}=0$.}
\begin{equation}
\label{eq:dfint}
\Delta^{\rm in}_f\, E_{nlj} \,=\,
 2 \delta_{\rm NP}(n,l) \,
\left\{
 \langle V^{(0)}_{\rm LS} \rangle_{nlj}
 + \langle V^{(0)}_{\rm T} \rangle_{nlj}
\right\} \,,
\end{equation}
where
\begin{equation}
\label{eq:dmpnl}
 \delta_{\rm NP}(n,l) \,=\,
 \frac{
\langle R^{(0)}_{nl} \vert
 r^{-3}
\vert R^{\rm NP}_{nl} \rangle}
{\langle R^{(0)}_{nl} \vert
 r^{-3}
\vert R^{(0)}_{nl} \rangle} \;;
\end{equation}
$R^{\rm NP}_{nl}$ is given in Eq.~(\ref{eq:wfnb}) and
$V_{\rm LS}^{(0)}$,
$V_{\rm T}^{(0)}$ are the leading (tree level) pieces of
$V_{\rm LS}$, $V_{\rm T}$. Using the explicit expressions for these
we have,
\begin{eqnarray}
\label{eq:expvlsz}
\langle V_{\rm LS}^{(0)}\rangle_{nlj} &=&
[j(j+1) - l(l+1) - 2]\,
\frac{3 C_F^4 \alpha_s {\widetilde{\alpha}_s}^3}
{16 n^3 l(l+1)(2 l +1)}\,m  \;;
\\
\label{eq:expvtz}
\langle V_{\rm T}^{(0)}\rangle_{nlj} &=&
\langle \frac{1}{2}\,S_{12} \rangle_{jl}
\,\,
\frac{C_F^4 \alpha_s {\widetilde{\alpha}_s}^3}
{8 n^2 l(l+1)(2 l +1)}\;m  \;;
\end{eqnarray}
with
\begin{equation}
\label{eq:sot}
\langle \frac{1}{2}\,S_{12} \rangle_{jl}
\, =
\,  \left\{
 \begin{array}{ll}
    -\frac{l+1}{2 l -1}  &,\, j=l-1 \\
     + 1                  &,\, j= l   \\
     -\frac{l}{2 l +3}  &, \, j= l+1 \,.
 \end{array}
  \right.
\end{equation}

The leading "external" fine structure shift,
$\Delta_f^{\rm ex}\,E_{nlj}$,
is caused by the crossed combination of the perturbations
\begin{displaymath}
- g  {\vec r} \cdot
{{\vec {\cal E}}\mkern -7mu{\lower1.6ex\hbox{$\widetilde {}$}}\;\,}
\;\;, \;\;
 \frac{g}{2 m^2} ({\vec S}\times {\vec p})  \cdot
{{\vec {\cal E}}\mkern -7mu{\lower1.6ex\hbox{$\widetilde {}$}}\;\,}
\;.
\end{displaymath}
In this case the external shift is also chromoelectric; the
chromomagnetic perturbation
$\displaystyle
 - \frac{g}{m} ({\vec S}_1 - {\vec S}_2)  \cdot
{{\vec {\cal B}}\mkern -7mu{\lower1.6ex\hbox{$\widetilde {}$}}\;\,}
$
does not contribute to the fine structure. The color algebra is now
like the one for the spin--independent shift, Subsection~4.1.
Thus,
\begin{eqnarray}
\nonumber
\Delta_f^{\rm ex} E_{nlj}
 &=& - 2 \frac{\pi {\langle \alpha_s G^2  \rangle}}{6 N_c}
\\
\label{eq:dfext}
&& \times \; \frac{1}{2 m^2} \,
 \sum_i
\,\left\langle \Psi^{(0)}_{nlj} \left\vert
 \,({\vec S}\times{\vec p})_i
\,\frac{1}{H^{\,'(0)}-E_n^{(0)}} \,r_i
\,\right\vert \Psi^{(0)}_{nlj} \right\rangle \;.
\end{eqnarray}
The angular momentum algebra, on the other hand, is
somewhat complicated. It is developed in detail in
Appendix~III for $n=2,\,l=1$. One gets
\begin{eqnarray}
\nonumber
\Delta_f^{\rm ex} E_{21j}
 &=& - \frac{\pi {\langle \alpha_s G^2  \rangle}}{6 N_c\,m^2}
 \left\{
 \frac{j(j+1)-4}{2}
\,\left\langle R^{(0)}_{21} \left\vert
\,\frac{1}{r}
\frac{1}{H_{2}^{\,'(0)}-E_2^{(0)}} \, r
\,\right\vert R^{(0)}_{21} \right\rangle
 \right.
\\
\nonumber
&& \left. \qquad + \, \nu(j)\,\,
\left\langle R^{(0)}_{21} \left\vert
\, \frac{\partial}{\partial r} \left(
\frac{1}{H_{0}^{\,'(0)}-E_2^{(0)}} \,
- \,\frac{1}{H_{2}^{\,'(0)}-E_2^{(0)}}
\right)  \,r \,
\right\vert R^{(0)}_{21} \right\rangle \,\right\}\,,
\\
\nonumber
&& \frac{1}{2}\nu(0) \,=\,
 \nu(1) \,=\, -\nu(2) \,=\, \frac{1}{3} \;.
\end{eqnarray}
The calculation is finished using the inverses of Appendix~I.
The result is
\begin{eqnarray}
\nonumber
\Delta_f^{\rm ex} E_{21j}
 &=&
\frac{1780\, [j(j+1)-4] - 2784\, \nu(j)}{9945}\,
\frac{\pi {\langle \alpha_s G^2  \rangle}}
{m^3 (C_F {\widetilde{\alpha}_s})^2}
\\
\label{eq:dfexta}
&\equiv&
 \,K(j)\,
 \frac{\pi {\langle \alpha_s G^2  \rangle}}
{m^3 (C_F {\widetilde{\alpha}_s})^2}
\;;
\end{eqnarray}
with
\begin{equation}
\label{eq:kj}
 K(0) \,=\, - \frac{8976}{9945} \;,\;
 K(1) \,=\, -K(2) \,=\, \frac{1}{2} \,K(0) \;.
\end{equation}
The perturbative fine splitting is (for $s=1$; the splitting
should be considered to vanish for $s=0$)
\begin{eqnarray}
\nonumber
\Delta^{\rm p}_{\rm f}\,E_{nlj} \,&=&\,
 \frac{3 C_F^4 \alpha_s(\mu^2)
 {\widetilde{\alpha}_s}^3(\mu^2)}{16 n^3 l(l+1)(2 l +1)}
\,m \,[j(j+1)-l(l+1)-2]\,
[1 + \delta_{\rm wf}(n,0)]^2
\\
\nonumber
&& \times
\left\{
 1 + \left[ \left(\frac{\beta_0}{2}-2 \right)
\left(
 \ln n -1 -\psi(n+l+1)+\psi(2 l+3)+\psi(2 l)
\right. \right. \right.
\\
\nonumber
&& \left. \left. \left.
\qquad
- \frac{n-l-1/2}{n}
 \right)
+ \frac{125 - 10\,n_f}{36} + \frac{\beta_0}{2}
\ln\frac{\mu}{m C_F {\widetilde{\alpha}_s}}
 + 2 \ln C_F {\widetilde{\alpha}_s}
\right]
\,\frac{\alpha_s}{\pi} \right\}
\\
\nonumber
&+&
 \frac{C_F^4 \alpha_s(\mu^2)
 {\widetilde{\alpha}_s}^3(\mu^2)}{8 n^3 l(l+1)(2 l +1)}
\,m \,
\langle \frac{1}{2}\,S_{12} \rangle_{lj}
\,[1 + \delta_{\rm wf}(n,0)]^2
\\
\nonumber
&& \times
\left\{
 1 + \left[ D + \left(\frac{\beta_0}{2}-3 \right)
\left(
 \ln n -\psi(n+l+1)+\psi(2 l+3)+\psi(2 l)
\right. \right. \right.
\\
\label{eq:dfp}
&& \left. \left. \left.
\qquad
- \frac{n-l-1/2}{n}
 \right)
+ \frac{\beta_0}{2}
\ln\frac{\mu}{m C_F {\widetilde{\alpha}_s}}
 + 3 \ln C_F {\widetilde{\alpha}_s}
\right]
\,\frac{\alpha_s}{\pi} \right\} \;.
\end{eqnarray}
The constants as in Eqs.~(\ref{eq:hls}), (\ref{eq:ht}) and
(\ref{eq:dwf}).

The full, relativistic plus radiative plus NP fine splitting
is then
\begin{equation}
\label{eq:dffull}
\Delta_{\rm f}\,E_{nlj} \,=\,
\Delta^{\rm p}_{\rm f}\,E_{nlj} +
\Delta^{\rm in}_{\rm f}\,E_{nlj} +
\Delta^{\rm ex}_{\rm f}\,E_{nlj} \;,
\end{equation}
the various terms given in Eqs.~(\ref{eq:dfint}),
(\ref{eq:dfexta})
and
(\ref{eq:dfp}).

\subsection{Decays into $e^+ e^-$}

For a state with $l=0$ the decay rate into $e^+ e^-$ is
given by
\begin{eqnarray}
\nonumber
 \Gamma(\Upsilon(n S) \rightarrow e^+ e^-)
&=& \frac{2}{n^3} \left[ \frac{Q_b \alpha}{M(\Upsilon(n S))}
 \right]^2\,
 [m C_F {\widetilde{\alpha}_s}(\mu^2)]^3
\\
\label{eq:decaynS}
&& \times \;
(1 + \delta_r)\,
 [1 + \delta_{\rm wf}(n,0) + \rho_{\rm NP}(n)]^2 \;.
\end{eqnarray}
Here $\delta_r$ is a "hard" radiative correction~\cite{bb:barb},
\begin{equation}
\label{eq:dhard}
 \delta_r \,=\, - \frac{4 C_F \alpha_s}{\pi} \;,
\end{equation}
$\delta_{\rm wf}(n,0)$ is given in Eq.~(\ref{eq:dwf}) and
$\rho_{\rm NP}(n)$ is the ratio of NP to unperturbed wave
functions at the origin:
\begin{equation}
\label{eq:ro}
 \rho_{\rm NP}(n) \,=\,
 \frac{ R^{\rm NP}_{n0}(0)}{ R^{(0)}_{n0}(0)} \;.
\end{equation}
It is to be calculated with the expressions of Appendix~II.

\section{Properties of Bottomium in States with $n=1,\,2$.}

We will use spectroscopic notation: states will be labeled
${n\,}^{2 s+1}l_j$, $l=0,\,1,\,2\,\ldots$ or
$S,\,P,\,D,\,\ldots$. The somewhat whimsical notation of the
Particle Data Tables (PDT)~\cite{bb:data} will also be indicated.
For $n=1,\,2,\,3$ mixing does not occur.

\subsection{States with $n=1$.}

{}From TY we have
\begin{eqnarray}
\nonumber
 M({1\,}^3\! S_1) &=& M(\Upsilon) \,=\,
 2 m \,\left\{
 1 - \frac{C_F^2 \,{\widetilde{\alpha}_s}^2(\mu^2)}{8}
 - \frac{C_F^2 \beta_0 \,\alpha_s^2(\mu^2)
\,{\widetilde{\alpha}_s}(\mu^2)}{8 \pi}
\right.
\\
\label{eq:mup}
&&
\left.
 \times \;\left( \ln\frac{\mu}{m C_F {\widetilde{\alpha}_s}} + 1
 - \gamma_{\rm E} \right) \right\}
 + \frac{\epsilon_{10} \pi
 {\langle \alpha_s G^2  \rangle}}{(m C_F {\widetilde{\alpha}_s})^4}
 \,m   \;,
\\
\nonumber
&& \epsilon_{10} \,=\, \frac{1872}{1275} \,\simeq\, 1.468 \;.
\end{eqnarray}
The order $\alpha_s^4$ is partially known\footnote{It includes
leading relativistic corrections
${\cal O}(\alpha_s^4)$,
one--loop radiative ones
${\cal O}((\alpha_s^4/\pi)\ln\mu^2)$ and
${\cal O}(\alpha_s^4/\pi)$, and leading logarithm two--loop
corrections
${\cal O}((\alpha_s^4/\pi^2)\ln^2\mu^2)$. The error
of Eq.~(\ref{eq:mhfone}) should be at the 10 to 20 $\%$ level.};
it adds to the right--hand--side of Eq.~(\ref{eq:mup}) a term
\begin{eqnarray}
\nonumber
&& 2 m\,
 \left[
 -\frac{3 C_F^4}{16} \left( 1 +
\left(a_1 + \frac{\gamma_{\rm E} \beta_0}{2} \right)
\frac{\alpha_s}{\pi} \right) \alpha_s
 {\widetilde{\alpha}_s}^3
+ \frac{C_F^3 a_2}{8}\,\alpha_s^2
 {\widetilde{\alpha}_s}^2
 \right.
\\
\label{eq:mupadd}
 &&
\phantom{2 m}
\, \left.
 - \frac{5 C_F^4}{128}  {\widetilde{\alpha}_s}^4
 - \frac{3 C_F^4 \beta_0^2}{16 \pi^2}
 \left( \ln\frac{\mu}{m C_F {\widetilde{\alpha}_s}}
 -1 -\gamma_{\rm E} \right) \alpha_s^3 {\widetilde{\alpha}_s}
 + \frac{C_F^4}{6} \alpha_s  {\widetilde{\alpha}_s}^3
 \right]
\;.
\end{eqnarray}
We will use both Eq.~(\ref{eq:mup}) alone and
Eqs.~(\ref{eq:mup}) plus (\ref{eq:mupadd}).

The hyperfine splitting is obtained from
Eq.~(\ref{eq:dhffull}), $\Delta^{\rm in}_{\rm hf}$
(Eq.~(\ref{eq:dhfinta})) evaluated with the expressions
for the $R$'s of Appendix~II, and the inverse in
Eq.~(\ref{eq:dhfext}) with those of Appendix~I. The result is
\begin{eqnarray}
\nonumber
 M({1\,}^3 S_1) - M({1\,}^1 S_0)
\,&=&\, M(\Upsilon) - M(\eta_b)
\\
\nonumber
&=& \,
 \frac{C_F^4 \alpha_s(\mu^2)
 {\widetilde{\alpha}_s}^3(\mu^2)}{3}\,m
\\
\nonumber
&&
\!\!\!\!
\!\!\!\!
\!\!\!\!
\!\!\!\!
\!\!\!\!
\times
\left\{
 1 + \left[ \frac{\beta_0}{2}\left(
 \ln\frac{\mu}{m C_F {\widetilde{\alpha}_s}}
+ \gamma_{\rm E}
 \right)
 -\frac{21}{4}\left(
 \ln\frac{1}{C_F {\widetilde{\alpha}_s}} -1 \right)
+ B \right]
\,\frac{\alpha_s}{\pi} \right\}
\\
\label{eq:mhfone}
&&
\!\!\! + \; \frac{C_F^4 \alpha_s(\mu^2)
 {\widetilde{\alpha}_s}^3(\mu^2)}{3}\,m
 \,\left[ \frac{270459}{108800} +
 \frac{1161}{8704} \right] \,
 \frac{\pi {\langle \alpha_s G^2  \rangle}}
 {m^4 {\widetilde{\alpha}_s}^6} \;.
\end{eqnarray}
In the NP contribution the first term is the internal,
the second the external which is, as is generally the case,
substantially smaller than the first. The term in square
brackets is, after multiplying by $\pi$,
$7.81 + 0.42 = 8.23$, slightly smaller
than the value given by Leutwyler~\cite{bb:leut} which was
also used in TY and equal to $10.2$. The difference in the value of
the hyperfine splitting, however, is fairly small. The
corrected value, following from Eq.~(\ref{eq:mhfone}), will be
given below.

For the $e^+ e^-$ decay Eq.~(\ref{eq:decaynS}) gives us
\begin{eqnarray}
\nonumber
\Gamma({1\,}^3 S_1 \rightarrow e^+ e^-)
 &=&
\Gamma(\Upsilon \rightarrow e^+ e^-)
\\
\nonumber
&=& 2 \left[ \frac{Q_b \alpha}{M(\Upsilon)} \right]^2 \,
 \left[ m C_F {\widetilde{\alpha}_s}(\mu^2) \right]^3
 \left( 1 - \frac{4 C_F \alpha_s}{\pi} \right)
\\
\label{eq:decay1S}
&&
\!\!\!\!\!
\!\!\!\!\!
\!\!\!\!\!
\times \;\,
 \left[
 1 + 3 \beta_0 \left(\ln\frac{2 \mu}{m C_F {\widetilde{\alpha}_s}}
 + \frac{1}{2} - \gamma_{\rm E} \right) \frac{\alpha_s}{4 \pi}
 + \frac{270459}{217600} \, \frac{\pi {\langle \alpha_s G^2  \rangle}}
 {m^4 {\widetilde{\alpha}_s}^6} \right]^2 \;,
\end{eqnarray}
and we have inserted the explicit values for $\delta_r$,
$\delta_{\rm wf}$, $\rho_{\rm NP}$.

\subsection{States with $n=2$. Spin--independent shifts.
Decay into $e^+ e^-$.}

We will denote by ${\overline M}({2\,}^3 P)$ the average of the
masses of the states\footnote{ Denoted by $\chi_{bj}(1 P)$ by the
PDT people, Ref.~\cite{bb:data}.} ${2\,}^3 P_j,\,j=0,\,1,\,2$:
\begin{equation}
\label{eq:massav}
 {\overline M}({2\,}^3 P)
\, =\, \frac{1}{9} \,\left\{
 5\, M({2\,}^3 P_2) + 3\,  M({2\,}^3 P_1) +  M({2\,}^3 P_0)
 \right\} \,=\, 9900 \pm 1 {\rm MeV} \;.
\end{equation}

{}From the analysis of TY and Ref.~\cite{bb:leut} we have,
\begin{eqnarray}
\nonumber
M({2\,}^3 S_1) - M({1\,}^3 S_1)
&=& M(\Upsilon(2 S)) - M(\Upsilon(1 S))
\\
\nonumber
&=&
2 m\, \left\{
\frac{3 C_F^2
{\widetilde{\alpha}_s}^2(\mu^2)}{32}
 \right.
\\
\nonumber
&& \quad
 \left.
+ \frac{C_F^2 \beta_0 \alpha_s
{\widetilde{\alpha}_s}}{32}
 \left[ 3 \ln\frac{\mu}{C_F m {\widetilde{\alpha}_s}}
+ \frac{5}{2} - 3 \gamma_{\rm E} -\ln 2 \right]
\frac{\alpha_s}{\pi} \right\}
\\
\label{eq:balms}
&& + m\, \frac{(2^6\,\epsilon_{20} - \epsilon_{10}) \pi
 {\langle \alpha_s G^2  \rangle} }{C_F^4 m^4
 {\widetilde{\alpha}_s}^4}  \; , \; \epsilon_{20} =
 \frac{2102}{1326} \simeq 1.585 \;;
\\
\nonumber
{\overline M}({2\,}^3 P) - M({1\,}^3 S_1)
 &=&
 2 m\, \left\{
\frac{3 C_F^2
{\widetilde{\alpha}_s}^2(\mu^2)}{32}
 \right.
\\
\nonumber
&& \quad
 \left.
+ \frac{C_F^2 \beta_0 \alpha_s
{\widetilde{\alpha}_s}}{32}
 \left[ 3 \ln\frac{\mu}{C_F m {\widetilde{\alpha}_s}}
+ \frac{13}{6} - 3 \gamma_{\rm E} -\ln 2 \right]
\frac{\alpha_s}{\pi} \right\}
\\
\label{eq:balmp}
&& + m\, \frac{(2^6\,\epsilon_{21} - \epsilon_{10}) \pi
 {\langle \alpha_s G^2  \rangle} }{C_F^4 m^4
 {\widetilde{\alpha}_s}^4}  \; , \; \epsilon_{21} =
 \frac{9929}{9945} \simeq 0.9984 \;.
\end{eqnarray}
It is interesting to consider on its own the
"Lamb shift", difference between Eqs.~(\ref{eq:balms}) and
(\ref{eq:balmp}), as here only the states with $n=2$ are
involved:
\begin{equation}
\label{eq:lamb}
M({2\,}^3 S_1) - {\overline M}({2\,}^3 P)
= 2 m\,
 \frac{C_F^2 \beta_0 \alpha_s^2
{\widetilde{\alpha}_s}}{96 \pi}
 + m\, \frac{2^6\,(\epsilon_{20} - \epsilon_{21})\, \pi
 {\langle \alpha_s G^2  \rangle} }{C_F^4 m^4
 {\widetilde{\alpha}_s}^4}  \;.
\end{equation}
As for the decay $\Upsilon(2 S) \rightarrow e^+ e^-$,
Eq.~(\ref{eq:decaynS}) gives
\begin{eqnarray}
\nonumber
 \Gamma({2\,}^3 S_1 \rightarrow e^+ e^-)
&=& \frac{1}{4} \left[ \frac{Q_b \alpha}{M(\Upsilon(2 S))}
 \right]^2\,
 [m C_F {\widetilde{\alpha}_s}(\mu^2)]^3
\\
\label{eq:decay2S}
\times \,
\left(1 - \frac{4 C_F \alpha_s}{\pi} \right)
\!\!\!
\!\!\!
\!\!
&&
\!\!
\!\!\!
\!\!\!
\, \left[
1 + 3 \beta_0 \left(\ln\frac{2 \mu}{m C_F
 {\widetilde{\alpha}_s}} + \frac{1}{2} -\gamma_{\rm E} \right)
 \frac{\alpha_s}{4 \pi}
 + \frac{302859}{884} \,
\frac{\pi {\langle \alpha_s G^2  \rangle}}
 {m^4 {\widetilde{\alpha}_s}^6}
 \right]^2 \;.
\end{eqnarray}

\subsection{States with $n=2$. Fine splittings.}

{}From Eq.~(\ref{eq:dffull}) and after some work we get the
fine structure splittings\footnote{Because $\delta_{\rm NP}$,
$\delta_{\rm wf}$ are large we have included them in a factor
$[1 + \delta_{\rm wf}]^2\,(1 + 2\delta_{\rm NP})$
in Eq.~(\ref{eq:mfj}). This form or the equivalent one of a
factor $[1 + \delta_{\rm wf}+\delta_{\rm NP}]^2$ are the
ones that give more {\em stable} numerical results.}
%\vskip.9cm
\begin{eqnarray}
\nonumber
&&
M({2\,}^3 P_j) -{\overline M}({2\,}^3 P)
\,=\,  m\,
C_F^4 \beta_0 \,\alpha_s(\mu^2)
\,{\widetilde{\alpha}_s}^3(\mu^2)
\\
\nonumber
&& \quad \times
\, \left[
1 + 3 \beta_0 \left(\ln\frac{2 \mu}{m C_F
 {\widetilde{\alpha}_s}} + \frac{5}{6} -\gamma_{\rm E} \right)
 \frac{\alpha_s}{4 \pi} \right]^2
\left(
 1 + \frac{111699}{221} \,
\frac{\pi {\langle \alpha_s G^2  \rangle}}
 {m^4 {\widetilde{\alpha}_s}^6}
 \right)
\\
\nonumber
&& \quad \times
 \left\{\; \frac{j(j+1)-4}{256}
\left\{ 1 + \,\left[\left( \frac{\beta_0}{2}-2 \right)
 \left( \ln\frac{2 \mu}{m C_F
 {\widetilde{\alpha}_s}} -\gamma_{\rm E} \right)
 + 2 \ln\frac{\mu}{m} + \frac{125 -10\,n_f}{36} \right]
 \frac{\alpha_s}{\pi} \right\} \right.
\\
\nonumber
&& \quad \quad\;\;
 \left.
 +  \,\frac{\langle \frac{1}{2} S_{12} \rangle_{1j}}{384}
\left\{ 1 + \,\left[\left( \frac{\beta_0}{2}-3 \right)
 \left( \ln\frac{2 \mu}{m C_F
 {\widetilde{\alpha}_s}} +1 -\gamma_{\rm E} \right)
 + 3 \ln\frac{\mu}{m} + D \right]
 \frac{\alpha_s}{\pi} \right\} \;\right\}
\\
\label{eq:mfj}
 && \qquad\qquad\qquad\qquad\quad\;\; + \,\,m\,
\frac{K(j) \,\pi\, {\langle \alpha_s G^2  \rangle}}
 {m^4 (C_F {\widetilde{\alpha}_s})^6} \;.
\end{eqnarray}
The first term containing
${\langle \alpha_s G^2  \rangle}$ is the "internal" NP shift
(corresponding to Eq.~(\ref{eq:dfint})); the last term is
the "external" piece, Eq.~(\ref{eq:dfexta}). The experimental
shifts are
\begin{eqnarray}
\nonumber
M({2\,}^3 P_2) -  M({2\,}^3 P_1) &=& 21 \pm 1 \,{\rm MeV}
\\
\nonumber
M({2\,}^3 P_1) -  M({2\,}^3 P_0) &=& 32 \pm 2 \,{\rm MeV}\,.
\end{eqnarray}

\subsection{Hyperfine splittings for states with $n=2,\,l=1$.}

The hyperfine splitting
${\overline M}({2\,}^3 P) -  M({2\,}^1 P_1)$ has
not been measured experimentally for bottomium. For
charmonium,
\begin{equation}
\label{eq:hfcharm}
{\overline M}_{c\bar c}({2\,}^3 P)
-  M_{c\bar c}({2\,}^1 P_1) \,=\, - 0.9 \pm 0.2 \,{\rm MeV}\,.
\end{equation}
The theoretical calculation has been displayed in Subsection~4.2.
After substituing the explicit expressions for the various pieces
we get
\begin{eqnarray}
\nonumber
{\overline M}({2\,}^3 P) - M({2\,}^1 P_1)
&=& m\, \left( \frac{\beta_0}{2} - \frac{21}{4} \right)
 \,\frac{C_F^4 \alpha_s^2 {\widetilde{\alpha}_s}^3}{288 \pi}
\\
\label{eq:mhf}
 &&
+ \,\,m\,
\frac{61\,\pi\, {\langle \alpha_s G^2  \rangle}}
 {117\, m^4 {\widetilde{\alpha}_s}^2} \;.
\end{eqnarray}
This effect is remarkable. The coefficient
$\displaystyle \frac{\beta_0}{2} - \frac{21}{4}$ is
{\em negative}; hence the perturbative and all {\em internal} NP
contributions (which are, however, subleading) will be negative.
On the other hand, the {\em external} NP correction is
{\em positive}. For the (relatively) light quarks $c\bar c$, the
perturbative piece dominates; but for $b\bar b$, because it
decreases like
$\alpha_s^5$, and the NP one grows like
$\alpha_s^{-2}$, the situation is reversed and we will get
\begin{displaymath}
{\overline M}_{b\bar b}({2\,}^3 P)
-  M_{b\bar b}({2\,}^1 P_1)
  > 0 \,.
\end{displaymath}
This is of importance for calculations based on phenomenological
potentials
(see e.~g. Refs~\cite{bb:gupt,bb:halz}), a matter that will
be discussed in a separate
publication.

\section{Numerical Results.}

The numerical results which correspond to the formulas
given in the previous sections are presented in Table.~I.
Before discussing them a few words have to be said
about the calculational procedure. The quantities
pertaining exclusively to $b\bar b$ in states with $n=1$
have been taken from TY with the only modification of
the hyperfine $\Upsilon - \eta_b$ mass difference where
we have incorporated the (minute) modification following
our corrected evaluation of the NP contribution. The criterion
adopted in TY to choose the renormalization point $\mu$, was to
require that radiative and NP contributions be equal in
absolute value. Most results were in fact little dependent
on the actual value of $\mu$ chosen. The reason is that, for
$n=1$ the quark mass (as a function of $M(\Upsilon)$ taken as input)
begins at order $\alpha_s^0$ and the first corrections are
${\cal O}(\alpha_s^2)$. For the decay $\Upsilon \rightarrow e^+ e^-$,
the leading contribution is order $\alpha_s^3$; finally the "Balmer"
mass differences $M(\Upsilon n S) - M(\Upsilon 1 S)$ start at order
$\alpha_s^2$. By contrast the Lamb shift
$M({2\,}^3 S_1)-{\overline M}({2\,}^3 P)$ starts at
${\cal O}(\alpha_s^3)$,
the fine splittings among ${2\,}^3 P_j$ states
begin at order $\alpha_s^4$ (as does the $n=1$ hyperfine splitting)
and, finally, the hyperfine splitting
${\overline M}({2\,}^3 P)-M({2\,}^1 P_1)$ is an effect of
${\cal O}(\alpha_s^5)$. This means that for all these quantities
the choice of $\mu$ is essential as small variations in $\mu$ get
amplified.
Because of this we have chosen to {\em fit} the value of $\mu$.
We have considered three possibilities: fit the two fine
splittings, and then the Lamb shift and Balmer splitting
$M({2\,}^3 S_1) - M({1\,}^3 S_1)$ come out as predictions;
include the Lamb effect in the fit; or fit all four processes.
We present results in all three cases; we consider the last
possibility to give the optimum calculation. A remarkable
fact that lends credence to our results is that the values of
$\mu$ obtained with the three methods, as well as with the
criterion of TY (for the Lamb shift and Balmer splitting that
was considered also there) are extremely close one to the other.

The values of $\Lambda$,
${\langle \alpha_s G^2  \rangle}$ were {\em not} fitted.
We chose, as already mentioned,
\begin{eqnarray}
 \Lambda(n_f=3,{\rm 2\, loops}) &=& 250 \phantom{a}^{+80}_{-70}
\, {\rm MeV} \,,
\nonumber
\\
\label{eq:paras2}
 {\langle \alpha_s G^2  \rangle}
&=& 0.042 \pm 0.020 \; {\rm GeV}^4 \;.
\end{eqnarray}
Because we take $M(\Upsilon)$ as input, we {\em deduce}
$m_b$ (and ${\bar m}_b({\bar m}_b^2)$). For the pole mass,
Eq.~(\ref{eq:paras2}) implies, according to the analysis in TY,
\begin{equation}
\label{eq:mpole}
 m_b = 4906 \phantom{a}^{+69}_{-51}(\Lambda)
\phantom{a}^{-4}_{+4}({\langle \alpha_s G^2  \rangle})
\;{\rm MeV}\,,
\end{equation}
the first variation in Eq.~(\ref{eq:mpole}) tied to the variation of
$\Lambda$ in Eq.~(\ref{eq:paras2}), the second tied to that
of the gluon condensate also in Eq.~(\ref{eq:paras2}).

The agreement between theory and experimental data is remarkable,
as is remarkable the stability of the predictions of the
(as yet unmeasured) hyperfine splittings. The deviations are of the
expected order of the higher corrections, ${\cal O}(\alpha_s) \sim
30 \%$. As drawbacks, however, let us mention the fact that
some of the NP corrections, notably the ratio $\delta_{\rm NP}$, do
actually exceed unity\footnote{A list of some radiative
and NP contributions is given in Table~II.}. This makes the
results of the fine splittings less impressive than what
they look at first sight. Nevertheless, the choice of $\mu$ as well as
the way to write our equations certainly
allow a control of the results.

The process $\Upsilon({2\,}^3 S_1) \rightarrow e^+ e^-$ merits a
special discussion. If we take the central value
$\mu = 976 \,{\rm MeV}$ (Table I, column(c)) and consider the
leading expression of the width, i.e., we neglect
radiative and NP corrections, we get
\begin{equation}
\label{eq:dec23S1}
\Gamma^{(0)} \,=\, \frac{1}{4}
 \left[ \frac{Q_b \alpha}{M(2 S)}\right]^2 C_F^3 m^3
{\widetilde{\alpha}_s}^3
 \,=\,  0.77\, {\rm keV}\;.
\end{equation}
This is the value reported in Table~I, and it compares
favorably with experiment. Unfortunately the corrections
involve the factors
\begin{displaymath}
 (1 + \delta_r) \,,\,
 \left( 1 + \delta_{\rm wf}(2,0) \right)^2 \,,\,
 \left( 1 + \rho_{\rm NP}(2) \right)^2
\end{displaymath}
(see Eq.~(\ref{eq:decay2S}) for the expressions for the $\delta$,
$\rho$) and one has
\begin{displaymath}
\delta_r \,=\, -0.61 \;,\;
\delta_{\rm wf} \,=\, - 0.53 \; , \;
\rho_{\rm NP} \,=\, 3.6 \;.
\end{displaymath}
The prediction then becomes meaningless since the corrections
are much larger than the nominally leading term,
Eq.~(\ref{eq:dec23S1}); although here, as it happens in the
$c\bar c$ case (see TY) this leading term yields a
reasonable evaluation, considered as an order of magnitude
estimate.

Taken all together, our results here as well as those of TY,
constitute a coherent description of the lowest lying states
of heavy quark systems, using only rigorously derived QCD
properties and without need to have recourse to
phenomenological potentials or adjustable parameters.

\newpage

\appendix{\large\bf\underline{Appendix~I.}}

We evaluate the inverses
\begin{displaymath}
\frac{1}{H^{\kappa}_{l} - E^{(0)}_n} \,
 \rho^\nu \, {\rm e}^{-\rho/2}
 \,\equiv\, p_\nu(\rho) {\rm e}^{-\rho/2}\;.
\end{displaymath}
Here
\begin{displaymath}
\rho \equiv \frac{2 r}{n a} \;,\;
 E_n^{(0)} \,=\, - \frac{1}{m a^2 n^2}
 \,=\, - m \frac{C_F^2 {\widetilde{\alpha}_s}^2}{4 n^2} \;,\;
 a \,= \, \frac{2}{m C_F {\widetilde{\alpha}_s}} \;,
\end{displaymath}
and
\begin{displaymath}
H_l^\kappa \,=\, -\frac{1}{m} \frac{1}{r^2}
\,\frac{\partial}{\partial r}
\left( r^2 \frac{\partial}{\partial r} \right) +
\frac{l(l+1)}{m r^2} + \frac{\kappa {\widetilde{\alpha}_s}}{r}
\,.
\end{displaymath}
For $\nu$ integer $p_\nu$ turns out to be a polynomial:
\begin{displaymath}
 p_\nu (\rho) \,=\,
 \,\sum_{j=0}^{\nu +1} \, c_j\, \rho^j  \;,
\end{displaymath}
and
\begin{eqnarray*}
 c_{\nu+1} &=&
\,\frac{C_F}{\kappa\,n +(\nu+2) C_F}
\,\frac{m n^2 a^2}{4} \;,
\\
c_{j-1} &=&
\,\frac{C_F}{\kappa\,n + j\, C_F}
 \,\left[ j(j+1) - l(l+1) \right] \,\,c_j \;,\;\;\;\;
 j=l,\,l\!+\!1,\,\ldots,\,\nu\!+ \!1 \;;
\\
 c_j &=& \,0 \;,\;\;\;\; j < l \;.
\end{eqnarray*}
When $H_l^\kappa = H_l^{\,'(0)}$ those equations give
a unique well--defined $p_\nu$.
For $H_l^\kappa = H_l^{(0)}$ one should replace $n$ by $n +\epsilon$.
Then $p_\nu$ contains a singular coefficient, proportional to
$1/\epsilon$. However, when evaluating
\begin{displaymath}
\frac{1}{H^{(0)}_{l} - E^{(0)}_n} \,P_{nl}\,\,
 \rho^\nu \, {\rm e}^{-\rho/2}
\end{displaymath}
with $P_{nl}$ the projector orthogonal to the solution of
\begin{displaymath}
\left ( H^{(0)}_{l} - E^{(0)}_n \right)
R^{(0)}_{nl}\,=\,0 \;,
\end{displaymath}
the singular term drops out and the limit
$\epsilon \rightarrow 0$ may be taken.

\newpage

\appendix{\large\bf\underline{Appendix~II.}}

Here we list some nonperturbative energy shifts and wave functions
(spin--independent).
We write
\begin{displaymath}
 E^{\rm NP}_{nl} \,=\,
\frac{\epsilon_{nl}\,n^6\,\pi {\langle \alpha_s G^2  \rangle}}
 {(m C_F {\widetilde{\alpha}_s})^4}\,m \;.
\end{displaymath}
Then,
\begin{displaymath}
\renewcommand{\arraystretch}{2.0}
\begin{array}{lcl}
 \epsilon_{10} \,=\,
\displaystyle{\frac{624}{425}}
& \phantom{aaaaa} &
\epsilon_{20} \,=\,\displaystyle{\frac{1\,051}{663}}
\\
 \epsilon_{21} \,=\,\displaystyle{\frac{9\,929}{9\,945}}
& \phantom{aaaaa} &
  \epsilon_{30} \,=\,\displaystyle{\frac{769\,456}{463\,239}}
\\
  \epsilon_{31} \,=\,
\displaystyle{\frac{11\,562\,272}{8\,492\,715}}
& \phantom{aaaaa} &
  \epsilon_{40} \,=\,
\displaystyle{\frac{101\,509}{60\,060}}
\\
\epsilon_{50} \,=\,
\displaystyle{\frac{443\,288\,368}{260\,175\,675}}
& \phantom{aaaaa} & \phantom{aaaaaaa}
\end{array}
\renewcommand{\arraystretch}{1.0}
\end{displaymath}
For the wave functions, and with
$\displaystyle \rho \equiv \frac{2\,r}{n\,a}$,
\begin{eqnarray*}
 R^{\rm NP}_{10} &=& \frac{\pi {\langle \alpha_s G^2  \rangle}}
 { m^4 \,(C_F {\widetilde{\alpha}_s})^6}
 \,\frac{2}{a^{3/2}}\,{\rm e}^{-\rho/2}
\\ &&\quad \times\,
 \left\{
 \frac{2\,968}{425}
 -\frac{104}{425}\,\rho^2
 -\frac{52}{1\,275}\,\rho^3
 -\frac{1}{225}\,\rho^4
\right\}
\\
 R^{\rm NP}_{20} &=& \frac{\pi {\langle \alpha_s G^2  \rangle}}
 { m^4 \,(C_F {\widetilde{\alpha}_s})^6}
 \,\frac{1}{\sqrt{2}\,a^{3/2}}\,
{\rm e}^{-\rho/2}
\\ &&\quad \times
 \left\{
 \frac{3\,828\,736}{1\,989}
 -\frac{1\,914\,368}{1\,989}\,\rho
 -\frac{134\,528}{1\,989}\,\rho^2
 +\frac{67\,264}{5\,967}\,\rho^3
 +\frac{736}{663}\,\rho^4
 +\frac{16}{153}\,\rho^5
\right\}
\\
 R^{\rm NP}_{21} &=& \frac{\pi {\langle \alpha_s G^2  \rangle}}
 { m^4 \,(C_F {\widetilde{\alpha}_s})^6}
 \,\frac{1}{\sqrt{4!}\,a^{3/2}}\,\rho\,{\rm e}^{-\rho/2}
\\ &&\quad \times
 \left\{
 \frac{3\,299\,840}{1\,989}
 -\frac{149\,888}{5\,967}\,\rho^2
 -\frac{5\,248}{1\,989}\,\rho^3
 -\frac{32}{153}\,\rho^4
\right\}
\\
 R^{\rm NP}_{30} &=& \frac{\pi {\langle \alpha_s G^2  \rangle}}
 { m^4 \,(C_F {\widetilde{\alpha}_s})^6}
 \,\frac{1}{\sqrt{3}\,a^{3/2}}
\,{\rm e}^{-\rho/2}
\\ &&\quad \times
 \left\{
 \frac{189\,965\,808}{5\,719}
 -\frac{189\,965\,808}{5\,719}\,\rho
 +\frac{24\,735\,864}{5\,719}\,\rho^2
 +\frac{3\,462\,552}{5\,719}\,\rho^3
 -\frac{1302}{43}\,\rho^4
 \right.
\\
 && \quad \quad
 \left.
 -\frac{3\,042}{1\,505}\,\rho^5
 -\frac{9}{43}\,\rho^6
\right\}
\\
 R^{\rm NP}_{31} &=& \frac{\pi {\langle \alpha_s G^2  \rangle}}
 { m^4 \,(C_F {\widetilde{\alpha}_s})^6}
 \,\frac{1}{\sqrt{6}\,a^{3/2}}
\,{\rm e}^{-\rho/2}
\\ &&\quad \times
 \left\{
 \frac{1\,325\,287\,104}{62\,909}\,\rho
 -\frac{331\,321\,776}{62\,909}\,\rho^2
 -\frac{124\,833\,216}{314\,545}\,\rho^3
 +\frac{49\,872}{1\,505}\,\rho^4
 \right.
\\
 && \quad \quad
 \left.
 +\frac{3\,672}{1\,505}\,\rho^5
 +\frac{9}{43}\,\rho^6
\right\}
\end{eqnarray*}
\vskip1.4cm
\begin{eqnarray*}
 R^{\rm NP}_{40} &=& \frac{\pi {\langle \alpha_s G^2  \rangle}}
 { m^4 \,(C_F {\widetilde{\alpha}_s})^6}
 \,\frac{1}{a^{3/2}}
\,{\rm e}^{-\rho/2}
\\ &&\quad \times
 \left\{
 \frac{5\,609\,365\,504}{45\,045}
 -\frac{2\,804\,682\,752}{15\,015} \,\rho
 +\frac{57\,706\,496}{1\,001}\,\rho^2
 -\frac{20\,160\,512}{15\,015}\,\rho^3
 \right.
\\
 && \quad \quad
 \left.
 -\frac{93\,551\,104}{135\,135}\,\rho^4
 +\frac{59\,392}{3\,861}\,\rho^5
 +\frac{256}{429}\,\rho^6
 +\frac{32}{351}\,\rho^7
\right\}
\\
 R^{\rm NP}_{50} &=& \frac{\pi {\langle \alpha_s G^2  \rangle}}
 { m^4 \,(C_F {\widetilde{\alpha}_s})^6}
 \,\frac{1}{\sqrt{5}\,a^{3/2}}
\,{\rm e}^{-\rho/2}
\\ &&\quad \times
 \left\{
+\frac{37\,087\,558\,150\,000}{31\,221\,081}
-\frac{74\,175\,116\,300\,000}{31\,221\,081}\,\rho
+\frac{35\,702\,282\,000\,000}{31\,221\,081}\,\rho^2
 \right.
\\
 && \quad \quad
-\frac{13\,695\,312\,550\,000}{93\,663\,243}\,\rho^3
-\frac{561\,983\,427\,500}{93\,663\,243}\,\rho^4
+\frac{138\,527\,387\,500}{93\,663\,243}\,\rho^5
\\
 && \quad \quad
 \left.
-\frac{4\,827500}{261\,873}\,\rho^6
-\frac{1\,250}{9\,699}\,\rho^7
-\frac{625}{6\,588}\,\rho^8
\right\}\;.
\end{eqnarray*}
For ease of reference we also give the first $R^{(0)}$'s
\begin{eqnarray*}
 R^{(0)}_{10}(r) &=& \frac{2}{a^{3/2}}\, {\rm e}^{-r/a}
\\
 R^{(0)}_{20}(r) &=& \frac{1}{\sqrt{2}\,a^{3/2}}\,
\left(1- \frac{r}{a} \right)\,{\rm e}^{-r/2a}
\\
 R^{(0)}_{21}(r) &=& \frac{1}{2\,\sqrt{6}\,a^{3/2}}\,
\frac{r}{a}\,{\rm e}^{-r/2a}
\end{eqnarray*}
For the ${\bar R}^{(0)}_{nl}$'s, replace $a$ by $b(n,l)$ given in
Eq.~(\ref{eq:bnl}).

\newpage

\appendix{\large\bf\underline{Appendix~III.}}

We evaluate the matrix element ($21$ stands for $nl$)
\begin{displaymath}
{\cal M} = \,\sum_i \,
 \left\langle \Psi^{(0)}_{21j} \left\vert
 \left( {\vec S}\times {\vec P}\right)_i \,\,
 \frac{1}{H^{\,'(0)} - E_2^{(0)}} \,\, r_i \,
\right\vert
\Psi^{(0)}_{21j}
 \right\rangle \;.
\end{displaymath}
It is convenient to use a Cartesian basis for the
spin--angular momentum piece of $\Psi^{(0)}_{21j}$ so that
\begin{displaymath}
 \phantom{aaaaaaaaaaaaaaaaaaaaaaa}
 \Psi_{21j}({\vec r}) =
 \sum_{ik} \,
\xi_{ik}^{(\alpha)}(j)
\, {\hat r}_i \,
{\chi_k\mkern -14mu
{\lower2.1ex\hbox{$\widetilde {}$}}\;\;\,}
 \,R^{(0)}_{21}(r)\; .
 \phantom{aaaaaaaaaaaaaaaaaaaaa}
({\rm III}.\,1)
\end{displaymath}
Here $ {\hat r}= {\vec r}/r$, the
${\chi_k\mkern -14mu
{\lower2.1ex\hbox{$\widetilde {}$}}\;\;\,}
$ are column spin~$1$ wave functions and the coefficients
$\xi_{ik}^{(\alpha)}(j)
$ are
\begin{eqnarray*}
 &&
\xi_{ik}^{(0)}(0)
 = \frac{1}{\sqrt{4 \,\pi}}\,\delta_{ik} \;\;,\;\;\;
\xi_{ik}^{(a)}(1)
 = \frac{3}{\sqrt{8 \,\pi}}\,\epsilon_{aik} \;\;,
\\
 &&
\xi_{ik}^{(ab)}(2)
 = \frac{3}{\sqrt{4 \,\pi}}\,
 \left\{
\delta_{ia}\,\delta_{kb}
- \frac{1}{3}\,
\delta_{ik}\,\delta_{ab}
 \right\} \;.
\end{eqnarray*}
The last expression valid for $a \neq b$. The indices
$0,\,a,\,ab$, collectively denoted by $\alpha$ in~(III.$\,1$)
give the (Cartesian) third component of total angular
momentum. The spin--angular momentum wave functions
\begin{displaymath}
{\xi^{(\alpha)}\mkern -23mu
{\lower2.1ex\hbox{$\widetilde {}$}}\;\;\;\;\,}(j)
 =
 \sum_{ik} \,
\xi_{ik}^{(\alpha)}(j)
\,\, {\hat r}_i \,
{\chi_k\mkern -14mu
{\lower2.1ex\hbox{$\widetilde {}$}}\;\;\;\,}
\end{displaymath}
form an orthonormal set:
\begin{displaymath}
\int \!\!d\Omega\,\,
{\xi^{(\alpha)}\mkern -23mu
{\lower2.1ex\hbox{$\widetilde {}$}}\;\;\;\;\,}(j)
 \,
{\xi^{(\beta)}\mkern -23mu
{\lower2.1ex\hbox{$\widetilde {}$}}\;\;\;\;\,}(j')
 \,= \, \delta_{j j'}\,\delta_{\alpha \beta}\;.
\end{displaymath}
We have
\begin{eqnarray*}
{\cal M} &=& \,\sum_a \,
 \left\langle R^{(0)}_{21}
\,
{\xi\mkern -6mu
{\lower2.1ex\hbox{$\widetilde {}$}}\;\,}(j)
\,
\left\vert
 \left( {\vec S}\times {\vec P}\right)_a \,\,
 \frac{1}{H^{\,'(0)} - E_2^{(0)}} \,\, r_a \,
\right\vert
R^{(0)}_{21}
\,
{\xi\mkern -6mu
{\lower2.1ex\hbox{$\widetilde {}$}}\;\,}(j)
\,
 \right\rangle
\\
&=& \,
\sum_{\vphantom{a}^{\scriptstyle ik\,i'k'}_{\scriptstyle \,\,abc}}
\,
 \left\langle R^{(0)}_{21}
\left\vert
\vphantom{\frac{1}{H}}
 \int \!\!d\Omega\,\, \xi_{i'k'}(j)
 \, {\hat r}_{i'} \,
{\chi_{k'}^\dagger\mkern -17.5mu
{\lower2.1ex\hbox{$\widetilde {}$}}\;\;\,\,}
\,\epsilon_{abc}\, S_b \,P_c
 \right. \right.
\\
&& \quad\quad\quad\quad \times \; \left. \left.
 \frac{1}{H^{\,'(0)} - E_2^{(0)}} \,\, r_a \,{\hat r}_i
 \,\xi_{ik}(j) \,
{\chi_k\mkern -14mu
{\lower2.1ex\hbox{$\widetilde {}$}}\;\;\,}\,
\right\vert
R^{(0)}_{21}
 \right\rangle \;.
\end{eqnarray*}
If we write identically
\begin{displaymath}
 {\hat r}_a {\hat r}_i \,=\,
 \left( {\hat r}_a {\hat r}_i - \frac{1}{3} \,\delta_{ai}
 \right)
+ \frac{1}{3}\,\delta_{ai} \;,
\end{displaymath}
then the first term corresponds to angular momentum $2$,
and the second to angular momentum zero.
Therefore, when acting on
the first we may replace
$H^{\,'(0)}$ by $H^{\,'(0)}_2$, and when acting on the second
$H^{\,'(0)}$ by
$H^{\,'(0)}_0$. Hence,
\vskip1cm
\begin{eqnarray*}
{\cal M} &=& \,
\sum_{\vphantom{a}^{\scriptstyle ik\,i'k'}_{\scriptstyle \,\,abc}}
\,
 \left\langle R^{(0)}_{21}
\left\vert
\vphantom{\frac{1}{H}}
 \int \!\!d\Omega\,\, \xi_{i'k'}
 \, {\hat r}_{i'} \,
{\chi_{k'}^\dagger\mkern -17.5mu
{\lower2.1ex\hbox{$\widetilde {}$}}\;\;\,\,}
\,\epsilon_{abc}\, S_b \,P_c
 \right. \right.
\\
&& \quad\quad\quad\quad\quad \times \; \left. \left.
  \left( {\hat r}_a {\hat r}_i - \frac{1}{3} \,\delta_{ai}
 \right)
\, \frac{1}{H^{\,'(0)}_2 - E_2^{(0)}} \,\,
 \,\xi_{ik} \,
{\chi_k\mkern -14mu
{\lower2.1ex\hbox{$\widetilde {}$}}\;\;\,}\,r\,
\right\vert
R^{(0)}_{21}
 \right\rangle
\\
 && \!\!\!\!\!\! + \;
\frac{1}{3}\,
\sum_{\vphantom{a}^{\scriptstyle ik\,i'k'}_{\scriptstyle \,\,abc}}
\,
 \left\langle R^{(0)}_{21}
\left\vert
\vphantom{\frac{1}{H}}
 \int \!\!d\Omega\,\, \xi_{i'k'}
 \, {\hat r}_{i'} \,
{\chi_{k'}^\dagger\mkern -17.5mu
{\lower2.1ex\hbox{$\widetilde {}$}}\;\;\,\,}
\,\epsilon_{abc}\, S_b \,P_c \,\delta_{ai}
\, \frac{1}{H^{\,'(0)}_0 - E_2^{(0)}} \,\,
 \,\xi_{ik} \,
{\chi_k\mkern -14mu
{\lower2.1ex\hbox{$\widetilde {}$}}\;\;\,}\,r\,
\right\vert
R^{(0)}_{21}
 \right\rangle \;.
\end{eqnarray*}
and, after straightforward substitutions and arrangements,
\begin{eqnarray*}
{\cal M} &=&
\sum_{ik\,i'k'}
\,
 \left\langle R^{(0)}_{21}
\left\vert
\vphantom{\frac{1}{H}}
 \int \!\!d\Omega\,\, \xi_{i'k'}
 \, {\hat r}_{i'} \,
{\chi_{k'}^\dagger\mkern -17.5mu
{\lower2.1ex\hbox{$\widetilde {}$}}\;\;\,\,}
\,\,{\vec S}\cdot {\vec L}\,\,r_i\,\frac{1}{r}\,
\, \frac{1}{H^{\,'(0)}_2 - E_2^{(0)}} \,\,
 \,\xi_{ik} \,
{\chi_k\mkern -14mu
{\lower2.1ex\hbox{$\widetilde {}$}}\;\;\,}\,r\,
\right\vert
R^{(0)}_{21}
 \right\rangle
\\
&&
\!\!
\!\!
\!\!
+ \; \frac{1}{3}\,
 \sum_{ik\,i'k'\,cs}
\,\left(
\delta_{is}\,\delta_{ck} -
\delta_{ik}\,\delta_{cs} \right) \,
 \left\langle R^{(0)}_{21}
\left\vert
\vphantom{\frac{1}{H}}
 \int \!\!d\Omega\,\, \xi_{i'k'}
 \, {\hat r}_{i'} \,
{\chi_{k'}^\dagger\mkern -17.5mu
{\lower2.1ex\hbox{$\widetilde {}$}}\;\;\,\,}
 \,\xi_{ik} \,{\hat r}_c \,
{\chi_s\mkern -14mu
{\lower2.1ex\hbox{$\widetilde {}$}}\;\;\,}\,\right.\right.
\\
&&
\quad\quad\quad\quad \times \; \left.\left.
\, \frac{\partial}{\partial r} \left(
\frac{1}{H_{0}^{\,'(0)}-E_2^{(0)}} \,
- \,\frac{1}{H_{2}^{\,'(0)}-E_2^{(0)}}
\right)  \,r \,
\right\vert R^{(0)}_{21} \right\rangle \;.
\end{eqnarray*}
The only noteworthy aspects of the derivation are first, that,
because $H^{\,'(0)}_l$ only acts on the radial variable, and the
${\hat r}_i$ only depend on the angular ones,
\begin{displaymath}
\frac{1}{H^{\,'(0)}_l - E_2^{(0)}} \,{\hat r}_i
 \,=\,
{\hat r}_i\,
\frac{1}{H^{\,'(0)}_l - E_2^{(0)}} \;,
\end{displaymath}
and, second, that for any $f(r)$,
\begin{displaymath}
 P_k \,f(r) \,=\, -i\,{\hat r}_k \,\frac{\partial f(r)}
{\partial r} \;.
\end{displaymath}
The calculation is readily finished. Because
\begin{displaymath}
 \sum_{ik} \,
\xi_{ik}(j)
\,\, {\hat r}_i \,
{\chi_k\mkern -14mu
{\lower2.1ex\hbox{$\widetilde {}$}}\;\;\,}
\end{displaymath}
corresponds to total angular momentum $j$,
\begin{displaymath}
 {\vec S}\cdot{\vec L}\,\,
 \sum_{ik} \,
\xi_{ik}(j)
\,\, {\hat r}_i \,
{\chi_k\mkern -14mu
{\lower2.1ex\hbox{$\widetilde {}$}}\;\;\,}
  \,=\,
\frac{j(j+1)-l(l+1)-s(s+1)}{2}\,\,
 \sum_{ik} \,
\xi_{ik}(j)
\,\, {\hat r}_i \,
{\chi_k\mkern -14mu
{\lower2.1ex\hbox{$\widetilde {}$}}\;\;\,} \;,
\end{displaymath}
with $l=s=1$. Defining also
\begin{eqnarray*}
&& \nu(j) \,=\, \frac{4\,\pi}{9} \,
 \sum_{ij} \,
\left(
\xi_{ik}(j)\xi_{ki}(j) -
\xi_{ii}(j)\xi_{kk}(j)
\right) \;,
\\
&& \frac{1}{2}\nu(0) \,=\,
 \nu(1) \,=\, -\nu(2) \,=\, \frac{1}{3} \;,
\end{eqnarray*}
we finally get
\begin{eqnarray*}
\nonumber
 {\cal M}
 &=&
 \frac{4-j(j+1)}{2}
\,\left\langle R^{(0)}_{21} \left\vert
\,\frac{1}{r} \,
\frac{1}{H_{2}^{\,'(0)}-E_2^{(0)}} \,\, r
\,\right\vert R^{(0)}_{21} \right\rangle
\\
\nonumber
&&\!\!\!\!\!\!\!  + \; \nu(j)\,\,
\left\langle R^{(0)}_{21} \left\vert
\, \frac{\partial}{\partial r} \left(
\frac{1}{H_{0}^{\,'(0)}-E_2^{(0)}} \,
- \,\frac{1}{H_{2}^{\,'(0)}-E_2^{(0)}}
\right)  \,r \,
\right\vert R^{(0)}_{21} \right\rangle \;.
\end{eqnarray*}

% {\widetilde{\alpha}_s}
% {\langle \alpha_s G^2  \rangle}
%
% okay
%
%

%
%
%%%%***start table1*****
\renewcommand{\arraystretch}{1.5}
\begin{table}[b]
\begin{center}
\begin{tabular}{|@{}c@{}|@{}c@{}|@{}c@{}|@{}c@{}|@{}c@{}|}
\hline
Quantity & (a) & (b) & (c)  & Experiment
\\
\hline
  $\mu\,\, {\mbox{\footnotesize\rm{(MeV)}}}$
& $\,\,990\phantom{a}^{+213}_{-198}\phantom{a}^{-43}_{+90}$
& $\,\,968\phantom{a}^{+231}_{-224}\phantom{a}^{-53}_{+104}$
& $\,\,976\phantom{a}^{+238}_{-228}\phantom{a}^{-54}_{+107}$
&
\\
\hline
  $\alpha_s(\mu^2)$
& $\,\, 0.36\pm 0.03 \phantom{a}^{+0.01}_{-0.02}\;$
& $\,\, 0.37\pm 0.02 \phantom{a}^{+0.01}_{-0.03}\;$
& $\,\, 0.36\phantom{a}^{+0.03}_{-0.01} \pm 0.02\;$
&
\\
\hline
  $ {\widetilde{\alpha}_s}(\mu^2)$
& $\,\, 0.54\phantom{a}^{+0.07}_{-0.06} \phantom{a}^{+0.02}_{-0.04}$
& $\,\, 0.55\phantom{a}^{+0.06}_{-0.03} \phantom{a}^{+0.03}_{-0.05}$
& $\,\, 0.55\phantom{a}^{+0.05}_{-0.03} \phantom{a}^{+0.03}_{-0.05}$
&
\\
\hline
  $ {2\,}^3 P_2 - {2\,}^3 P_1  $
& $\,\, 22.2 \phantom{a}^{-0.4}_{+0.2} \pm 0$
& $\,\, 18.9 \phantom{a}^{+2.3}_{-6.1} \phantom{a}^{-2.6}_{+2.5}$
& $\,\, 20.0 \phantom{a}^{+1.2}_{-6.7} \phantom{a}^{-2.5}_{+2.6}$
& $\,\, 21 \pm 1 \,\, {\mbox{\footnotesize\rm{MeV}}}$
\\
\hline
  $ {2\,}^3 P_1 - {2\,}^3 P_0  $
& $\,\, 30.0 \pm 0.6 \phantom{a}^{+0}_{-0.2}$
& $\,\, 25.6  \phantom{a}^{+4.2}_{-9.0} \phantom{a}^{-3.6}_{+3.1}$
& $\,\, 27.2 \phantom{a}^{+5.2}_{-9.8} \phantom{a}^{-3.6}_{+3.2}$
& $\,\, 32 \pm 2 \,\, {\mbox{\footnotesize\rm{MeV}}}$
\\
\hline
  $ {2\,}^3 S_1 - \overline{{2\,}^3 P}  $
& $\,\, 193 \phantom{a}^{-49}_{+82} \phantom{a}^{+92}_{-50}$
& $\,\, 183 \mp 40 \phantom{a}^{+31}_{-42}$
& $\,\, 186 \phantom{a}^{-39}_{+40} \phantom{a}^{+32}_{-42}$
& $\,\, 123 \pm 1 \,\, {\mbox{\footnotesize\rm{MeV}}}$
\\
\hline
  $ {2\,}^3 S_1 - {1\,}^3 S_1  $
& $\,\, 487 \phantom{a}^{-148}_{+222} \phantom{a}^{+68}_{-69}$
& $\,\, 436 \phantom{a}^{-105}_{+74} \phantom{a}^{+14}_{-28}$
& $\,\, 455 \phantom{a}^{-97}_{+68} \phantom{a}^{+17}_{-32}$
& $\,\, 563 \pm 0.4 \,\, {\mbox{\footnotesize\rm{MeV}}}$
\\
\hline
  $ \overline{{2\,}^3 P} - {2\,}^1 P_1 $
& $\,\, 1.7 \phantom{a}^{-0.6}_{+0.7} \pm 0.7$
& $\,\, 1.6 \phantom{a}^{-0.5}_{+0.4} \pm 0.6$
& $\,\, 1.7 \phantom{a}^{-0.6}_{+0.4} \phantom{a}^{+0.5}_{-0.6}
\,\,{\mbox{\footnotesize\rm{MeV}}}$
&
\\
\hline
  $ \,{2\,}^3 S_1 \rightarrow e^+ e^- \,$
&
&
& $ \sim \,\,0.77 $
& $\,\, 0.56 \pm 0.10 \,\, {\mbox{\footnotesize\rm{MeV}}}$
\\
\hline
\hline
  $ {\overline{m}_b}({\overline{m}_b}^2)$
& \multicolumn{3}{@{}c@{}|}{
 $ 4397 \phantom{a}^{+7}_{-2}\phantom{a}^{-3}_{+4}
   \; {\mbox{\footnotesize\rm{MeV}}} \;{\rm (d)}$ }
& $ 4250 \pm 100 \;{\rm (f)}$
\\
\hline
  $ {1\,}^3 S_1 - {1\,}^1 S_0  $
& \multicolumn{3}{@{}c@{}|}{
 $ 33 \phantom{a}^{+13}_{-7}\phantom{a}^{+2}_{-5}
   \; {\mbox{\footnotesize\rm{MeV}}} \;{\rm (e)}$ }
&
\\
\hline
  $ {1\,}^3 S_1 \rightarrow e^+ e^-  $
& \multicolumn{3}{@{}c@{}|}{
 $ 1.01 \pm 0.02\phantom{a}^{+0.18}_{-0.24}
   \; {\mbox{\footnotesize{\rm keV}}}$ }
& $ 1.34 \pm 0.04$
\\
\hline
  $ {1\,}^3 S_1 \rightarrow 2\,\gamma  $
& \multicolumn{3}{@{}c@{}|}{
 $ 0.17
   \; {\mbox{\footnotesize\rm{keV}}} \;{\rm (d)}$ }
&
\\
\hline
\end{tabular}
\end{center}
\caption{Compilation of results.}
\vskip0.6cm
\vbox{
\indent
Theoretical predictions, and experimental values for
$b\bar b$ states with $n=2,\,1$ and $l=1,\,0$,
$s=1,\,0$, $j=0,\,1,\,2$.
\begin{itemize}
 \item[(a)] The parameter $\mu$ obtained by fitting
 ${2\,}^3P_j$.
 \item[(b)] Fit including also
 $ {2\,}^3 S_1 - \overline{{2\,}^3 P}  $.
 \item[(c)] Fit with the former and
 ${2\,}^3 S_1 - {1\,}^3 S_1$.
$\left(\chi^2/{\rm degrees\, of\, freedom} =
 \left( 0.33
\phantom{a}^{-0.15}_{+0.62}
\phantom{a}^{+0.39}_{-0.23}
 \right)/3 \right)
$.
 \item[(d)] Result from TY.
 \item[(e)] Result with analysis from TY with corrected
 NP contribution
(N.~B.: old result, $35$ MeV).
 \item[(f)] Values obtained from
$ e^+ e^- \rightarrow \;{\rm hadrons}$
via QCD sum rules, see Refs.~4.
\end{itemize}
}
\end{table}
\renewcommand{\arraystretch}{1.0}
%%%**end table1****
%
%
\newpage
%%%%***start table2*****
\renewcommand{\arraystretch}{1.8}
\begin{table}[b]
\begin{center}
\begin{tabular}{|c|c|c|c|c|c|c|}
\hline
Quantity & tree (a) & tree + rad. (b) & NP ext. (c)  & $\;\delta_{\rm wf}\;$
 & $\;\delta_{\rm NP}\;$ & Total
\\
\hline
  $ {2\,}^3 P_2 - {2\,}^3 P_1  $
& $\;11.5\;$
& $\;3.4\;$
& $\;1.9\;$
& $\;-0.27\;$
& $\;2.2\;$
& $\;20.0\;$
\\
\hline
  $ {2\,}^3 P_1 - {2\,}^3 P_0  $
& $\;14.4\;$
& $\;4.9\;$
& $\;0.95\;$
& $\;-0.27\;$
& $\;2.2\;$
& $\;27.0\;$
\\
\hline
\end{tabular}
\end{center}
\caption{  Sample set of contributions.}
\vskip0.2cm
\vbox{
\indent
\begin{displaymath}
{\rm with}\;\;
 \mu \,=\, 976\,{\rm MeV} \;;\;
 \Lambda(n_f=3,\,2\,{\rm loops}) \,=\, 250\,{\rm MeV} \;;\;
 {\langle \alpha_s G^2  \rangle} \,=\, 0.042\,{\rm GeV}^4 \;.
\end{displaymath}
\begin{itemize}
 \item[(a)] With tree level potential (including
relativistic corrections).
 \item[(b)]
 One loop radiative corrections.
 \item[(c)] External NP corrections.
\end{itemize}
\indent All dimensional numbers in MeV.
}
\end{table}
\renewcommand{\arraystretch}{1.0}
%%%**end table2****

\end{document}